# Distributional stability and deterministic equilibrium selection under heterogeneous evolutionary dynamics

Dai Zusai[*]

March 2, 2018

Very preliminary[1]

**Abstract**

In the presence of persistent payoff heterogeneity, the evolution of the aggregate strategy hugely depends on the underlying strategy composition under many evolutionary dynamics, while the aggregate dynamic under the standard BRD reduces to a homogenized smooth BRD, where persistent payoff heterogeneity averages to homogeneous transitory payoff shocks. In this paper, we consider deterministic evolutionary dynamics in heterogeneous population and develop the stronger concept of local stability by imposing robustness to persistent payoff heterogeneity. It is known that nonaggregability holds generically if the switching rate in a given evolutionary dynamic correlates with the payoff gain from a switch. To parameterize the payoff sensitivity of an evolutionary dynamic, we propose to use tempered best response dynamics with bounded support of switching costs.

---

[*]Department of Economics, Temple University, 1301 Cecil B. Moore Ave., RA 873 (004-04), Philadelphia, PA 19122, U.S.A. Tel:+1-215-204-8880. E-mail: zusai@temple.edu.

[1]For the most recent version, visit http://sites.temple.edu/zusai/research/DistStbl/.

# 1 Introduction

In this paper, we consider deterministic evolutionary dynamics in heterogeneous population and develop the stronger concept of local stability by imposing robustness to persistent payoff heterogeneity. Ely and Sandholm (2005) verify that the *aggregate strategy*—the (marginal) distribution of actions over the entire society—under the standard best response dynamic (standard BRD) can average to the smooth BRD where every agent receives an idiosyncratic shock in payoffs: that is, persistent payoff heterogeneity resolves to homogeneous transitory payoff shocks and the smoothed BRD is obtained as a result of homogenization. However, Zusai (2018a) finds that it is not the case for many of other evolutionary dynamics; as long as switching rates to new actions are correlated with the payoff gains from switches, a heterogeneous dynamic cannot reduce to a homogenized dynamic; there is distributional disturbance from the average homogenized dynamic.[2] So, he proposes to look directly at strategy composition—the joint distribution of types and actions; at this level, he verifies that stationarity of Nash equilibrium in general and stability in potential games can be extended to those of equilibrium compositions in the heterogeneous setting.

Here we utilize this nonaggregability to narrow down the set of stable equilibria in a binary-choice game. For this, we compare the standard BRD and the tempered BRD (Zusai, 2018b): while the switching rate to the best response is always constant in the standard BRD, it increases with the payoff gain from the switch in the tempered BRD. Zusai (2018a) proves for potential games that the set of locally stable equilibria are the same in these two dynamics (and any "admissible" dynamics) as the set of local potential maximizers.[3] Combined it with the aggregability theorem by Ely and Sandholm (2005), we can infer that local stability of an equilibrium composition under these dynamics can be identified from local stability of the corresponding aggregate equilibrium under the homogenized smooth BRD. In particular for the standard BRD, the aggregability tells that the social state converges to a local stable equilibrium as long as the aggregate strategy belongs to a basin of attraction under the smooth BRD, regardless of the underlying strategy composition. However, it is not the case for local stability under nonaggregable dynamics such as the tempered BRD. Even if an equilibrium composition is locally stable and the initial aggregate strategy is arbitrarily close to–or even coincides with—the corresponding aggregate equilibrium, nonaggregable dynamics may drive the social state away from this equilibrium.

Dependency of the long-run outcome under nonaggregable dynamics on the strategy composition leads to a stronger stability concept of aggregate equilibrium by imposing robustness to distributional disturbance on the aggregate dynamic. We refine local stability of aggregate equilibrium by requiring it to attract any neighbor aggregate strategy, *regardless of the underlying strategy composition*; we call this refined stability *distributional stability*.

---

[2]The word "distributional" comes from distributional strategy, proposed by Milgrom and Weber (1985) to define rigorously the strategy composition in a Bayesian game.

[3]Admissibility requires only that an agent does not switch the action if the agent's current action is a best response to the current payoff vector (best response stationarity) and that the expected change in an agent's payoff from revision is positive whenever the agent's current action is not a best response (positive correlation of strategy revision and current payoffs).

By making the switching rate more sensitive to with the payoff gain of switch, we can raise the hurdle for an aggregate equilibrium to be selected; we can eventually select one equilibrium. The tBRD is useful for this equilibrium selection; we will see that the payoff sensitivity of the dynamic can be parameterized and also the identification of the most robust equilibrium reduces to payoff comparison between a few canonical types.

The paper proceeds as follows. In Section 2, we set up the model and the dynamic and review the fundamental facts about heterogeneous dynamics, which are studied in Zusai (2018a). In Section 3, distributional stability is defined. To characterize the basin of attraction that is robust to distributional disturbance, we introduce the concept of distributional critical masses. In Section 4, we make detailed analysis on the dynamic from a distributionally *instable* equilibrium. In Section 5, we propose equilibrium selection by distributional stability. The last section makes concluding remarks and discuss the possible link with global games and stochastic evolution.

## 2 Model

### 2.1 Aggregate games with payoff heterogeneity

We consider a large population of agents (the society) $\Omega := [0,1] \subset \mathbb{R}$ who share the same binary action set $\mathcal{A} = \{I, O\}$. We call those who choose $I$ (In) **participants**, imagining an entry game. We define probability measure $\mathbb{P}_\Omega : \mathcal{B}_\Omega \to [0,1]$ over the agent population as Lebesgue measure so $\mathbb{P}_\Omega(\Omega) = 1$. Denote by $\mathcal{B}_\Omega$ the Lebesgue $\sigma$-field over $\Omega$. Denote by $\mathfrak{a}(\omega) \in \{O, I\}$ the action taken by agent $\omega \in \Omega$. We restrict action profile $\mathfrak{a} : \Omega \to \mathcal{A}$ to a $\mathcal{B}_\Omega$-measurable function. Then, $\bar{x} := \mathbb{P}_\Omega(\{\omega \in \Omega : \mathfrak{a}(\omega) = I\}) \in [0,1]$ is the aggregate mass of participants in the entire population. We call it the **aggregate strategy**.

We focus on a binary **aggregate game** with additively separable payoff heterogeneity, as follows. The payoff from action $I$ depends only on the aggregate strategy $\bar{x}$ but it is common to all agents. So the payoff from $I$ at aggregate strategy $\bar{x} \in [0,1]$ is represented as $F(\bar{x}) \in \mathbb{R}$; let the payoff function $F : [0,1] \to \mathbb{R}$ be a Lipschitz continuous function. On the other hand, the payoff from action $O$ differs among agents but it does not change with the aggregate strategy or the action profile. So the payoff from $O$ for agent $\omega \in \Omega$ is represented as $\theta(\omega) \in \mathbb{R}$. The game exhibits **positive externality** if $F$ is an increasing function.

We call $\theta(\omega)$ the **type** of agent $\omega$. Agents' type profile $\theta : \Omega \to \mathbb{R}$ is assumed to be measurable with respect to $\mathcal{B}_\Omega$. Then, it induces probability measure $\mathbb{P}_\Theta : \mathcal{B}_\Theta \to [0,1]$ by $\mathbb{P}_\Theta(B_\Theta) := \mathbb{P}_\Omega(\theta^{-1}(B_\Theta))$ for each $B_\Theta \in \mathcal{B}_\Theta$. Denote by $\Theta \subset \mathbb{R}$ the support of $\mathbb{P}_\Theta$; we call it the type space. For this paper, we assume that $\mathbb{P}_\Theta$ is a continuous distribution over $\Theta$. Let $P_\Theta$ be the cumulative distribution function of $\mathbb{P}_\Theta$: i.e., $P_\Theta(\theta) = \mathbb{P}_\Theta((-\infty, \theta])$. Let $P_\Theta$ be the cumulative distribution function, i.e., $P_\Theta(\bar{\theta}) = \mathbb{P}_\Theta((-\infty, \bar{\theta}])$.

In contrast to the aggregate strategy, we define a **Bayesian strategy** $x : \Theta \to [0,1]$ to represent the participation rate of each type $\theta \in \Theta$. To define it rigorously, we first define **strategy composition** $X : \mathcal{B}_\Theta \to [0,1]$ as the distribution of participants over different types such that the marginal distribution of types coincides with $\mathbb{P}_\Theta$. Then, Bayesian strategy $x : \Theta \to [0,1]$ is defined as its Radon-Nikodym density such as $X(B_\Theta) = \int_{B_\Theta} x d\mathbb{P}_\Theta$ for each $B_\Theta \in \mathcal{B}_\Theta$. The aggregate strategy is expressed in terms of the Bayesian strategy as[4]

$$\bar{x} = \mathbb{E}_\Theta x.$$

Denote by $\mathcal{F}_\mathcal{X}$ the set of Bayesian strategies.[5] Similarly, let $\mathcal{X}$ be the set of strategy compositions. Due to one-to-one correspondence between $X$ and $x$, they can be seen as equivalent. We look at Bayesian strategies for simplicity in expositions and appealing to intuition.

Each agent's best response is determined from the aggregate strategy and its type: given the aggregate strategy $\bar{x} \in [0,1]$, a type-$\theta$ agent should take $I$ if $F(\bar{x}) > \theta$ and $O$ if $F(\bar{x}) < \theta$. The two actions $I$ and $O$ are indifferent for an agent with type $\theta = F(\bar{x})$; we call such a type the **indifferent type** given $\bar{x}$. Thanks to the assumption of a continuous type distribution, we can uniquely determine the Bayesian strategy and the aggregate strategy at the best response despite indeterminacy of the best response actions for agents of the indifferent type: when all the agents take the best responses, the Bayesian best response and the aggregate best response should be[6]

$$x^{\text{BR}}[\bar{x}](\theta) := \mathbf{1}\{\theta \leq F(\bar{x})\} = \begin{cases} 1 & \text{if } \theta \leq F(\bar{x}) \\ 0 & \text{if } \theta > F(\bar{x}), \end{cases} \quad \text{and} \quad \mathbb{E}_\Theta x^{\text{BR}}[\bar{x}] = P_\Theta(F(\bar{x})).$$

In a Nash equilibrium, (almost) every agent correctly predicts the strategy composition and takes the best response to it. Correspondingly, Bayesian strategy $x \in \mathcal{F}_\mathcal{X}$ is in **Bayesian equilibrium**, if $x = x^{\text{BR}}[\mathbb{E}_\Theta x]$, i.e.,

$$x(\theta) = x^{\text{BR}}[\mathbb{E}_\Theta x](\theta) = \mathbf{1}\{\theta \leq F(\bar{x})\} \qquad \text{for } \mathbb{P}_\Theta\text{-almost all } \theta \in \Theta. \tag{1}$$

Similarly, we define an **aggregate equilibrium** as an aggregate strategy $\bar{x}$ that satisfies

$$\bar{x} = \mathbb{E}_\Theta x^{\text{BR}}[\bar{x}] = P_\Theta(F(\bar{x})) \tag{2}$$

Notice that aggregate equilibrium does *not* imply that the underlying Bayesian strategy is a Bayesian equilibrium. Bayesian equilibrium needs complete sorting of agents by types, while only the total mass of participants matters to aggregate equilibrium.

## 2.2 Construction of heterogeneous evolutionary dynamics

In an evolutionary dynamic, an agent occasionally receives an opportunity to revise an action. The revision opportunity follows a Poison process, and here we assume that the arrival rate is one. While different evolutionary dynamics can be defined by different revision protocols that specify

---

[4]Here $\mathbb{E}_\Theta$ is the expectation operator on the probability space $(\Theta, \mathcal{B}_\Theta, \mathbb{P}_\Theta)$, while $\mathbb{E}_\Omega$ is that on $(\Omega, \mathcal{B}_\Omega, \mathbb{P}_\Omega)$: i.e., $\mathbb{E}_\Omega f := \int_\Omega f(\omega) \mathbb{P}_\Omega(d\omega)$ for a $\mathcal{B}_\Omega$-measurable function $f : \Omega \to \mathbb{R}$ and $\mathbb{E}_\Theta \tilde{f} := \int_\Theta \tilde{f}(\theta) \mathbb{P}_\Theta(d\theta)$ for a $\mathcal{B}_\Theta$-measurable function $\tilde{f} : \Theta \to \mathbb{R}$. If $f = \tilde{f} \circ \theta$, then we have $\mathbb{E}_\Theta \tilde{f} = \mathbb{E}_\Omega f$.

[5]Two Bayesian strategies $x, x' \in \mathcal{F}_\mathcal{X}$ are considered as identical, i.e., $x = x'$ if $x(\theta) = x'(\theta)$ for $\mathbb{P}_\Theta$-almost all $\theta \in \Theta$. They indeed yield the same strategy composition.

[6]Given condition $C(\theta)$ on type $\theta$, the indicator function $\mathbf{1}(S(\cdot)) : \Theta \to \{0,1\}$ returns the truth value of statement $C(\theta)$: $\mathbf{1}(C(\theta)) = 1$ if $C(\theta)$ is true, and $\mathbf{1}(C(\theta)) = 0$ if not. So, $\mathbb{E}_\Theta[\mathbf{1}(C(\theta))x_{a,0}(\theta)]$ is the mass of action-$a$ players whose types satisfy condition $C$.

which action to switch upon the receipt of a revision opportunity, we normally assume that an agent switches only to a better action. In a binary-choice game, "better" must be the "best." Thus, it is not restrictive to limit our attention to the following class of dynamics:

**Definition 1** (Exact optimization dynamics). Given payoff vector $\pi \in \mathbb{R}^2$, an agent who receives a revision opportunity switches from current action $i \in \mathcal{A}$ to another action $j \neq i$ with probability (switching rate)

$$\rho_{ij}(\pi(\theta)) = \begin{cases} 0 & \text{if } j \notin \operatorname{argmax}_{a \in \mathcal{A}} \pi_a(\theta), \\ Q_{ij}(\pi(\theta)) & \text{if } \{j\} = \operatorname{argmax}_{a \in \mathcal{A}} \pi_a. \end{cases}$$

Here, the conditional switching rate function $Q_{ij} : \mathbb{R}^{\mathcal{A}} \to \mathbb{R}_+$ is assumed to be a Lipschitz continuous function.

Then, the population dynamic is constructed as

$$\dot{x}(\theta) = v^F[x](\theta) := (1 - x(\theta))\rho_{OI}(F(\mathbb{E}_\Theta x), \theta) - x(\theta)\rho_{IO}(F(\mathbb{E}_\Theta x), \theta)$$

$$= \begin{cases} (1 - x(\theta))Q_{OI}(F(\mathbb{E}_\Theta x), \theta) & \text{if } I \text{ is optimal for } \theta, \text{ i.e., } F(\mathbb{E}_\Theta x) > \theta \\ -x(\theta)Q_{IO}(F(\mathbb{E}_\Theta x), \theta) & \text{if } O \text{ is optimal for } \theta, \text{ i.e., } F(\mathbb{E}_\Theta x) < \theta. \end{cases}$$

The canonical example of exact optimization dynamics is the (standard) BRD, formally defined in Gilboa and Matsui (1991) and Hofbauer (1995).

**Definition 2** (Standard best response dynamic; Gilboa and Matsui (1991); Hofbauer (1995)). The standard best response dynamic is defined as an exact optimization dynamic with $Q_{ij} \equiv 1$.

That is, upon the receipt of a revision opportunity, an agent switches to the best response to the current state with sure. With constant arrival rate 1 of a revision opportunity, it implies that agents switch to the best response at constant switching rate 1. Under the standard BRD in a heterogeneous setting, Bayesian strategy changes towards the Bayesian best response (given the current aggregate strategy) from the current Bayesian strategy at a constant speed 1.

$$\dot{x}(\theta) = x_{BR}[\mathbb{E}_\Theta x](\theta) - x(\theta) \qquad \text{for each } \theta \in \Theta.$$

Ely and Sandholm (2005) rigorously verify that the aggregate strategy $\bar{x} = \mathbb{E}_\Theta x$ under the heterogeneous standard BRD follows the **homogeneized smooth BRD** such that

$$\dot{\bar{x}} = P_\Theta(F(\bar{x})) - \bar{x},$$

in which the idiosyncratic payoff heterogeneity is only transient (independently drawn from the same $\mathbb{P}$ at each revision opportunity ) and the agent then switches to the ex-post optimal strategy with a constant rate. In particular, if the type distribution $\mathbb{P}_\Theta$ is a double exponential distribution, then the homogenized smooth BRD is a logit dynamic. The core implication of this homogenization is that, under the standard BRD, a strategy composition does not affect the dynamic of aggregate strategy. For example, as long as $\bar{x}$ is an aggregate equilibrium, then $\bar{x}$ must stay there even if $x$ is not a Bayesian equilibrium and thus agents are still switching actions.

In the presence of payoff heterogeneity, a constant switching rate sounds strong, since we can expect an agent who will gain a greater payoff improvement from a switch to be more likely to switch the action. This leads us to the tempered BRD (tBRD), proposed by Zusai (2018b):

**Definition 3** (Tempered best response dynamic; Zusai (2018b)). A tempered best response dynamic is defined as an exact optimization dynamic with $Q_{ij}(\boldsymbol{\pi}) = Q(\pi_j - \pi_i)$. For a *tempering* function $Q : \mathbb{R} \to [0,1]$, assume that $Q(q) = 0$ for any $q \leq 0$, $Q'(q) > 0$ for any $q \in (0, \bar{q})$ with some $\bar{q} \in \bar{\mathbb{R}}_+ := \mathbb{R}_+ \cup \{+\infty\}$ and $Q'(q) = 0$ for any $q \in [\bar{q}, +\infty)$.

Here, $\pi_j - \pi_i$ is the payoff improvement from switching action from $i$ to $j$. If $j$ is a best response, it is just equal to the **payoff deficit** of the current action $i$, i.e., $\max_{j \in \mathcal{A}} \pi_j - \pi_i$. In the tBRD, the switching rate of an agent increases with the agent's payoff deficit. Thus, payoff heterogeneity causes difference in switching rates over different types of agents through this payoff-dependent tempering function $Q$.

Note that the tempered BRD can be constructed by introducing a stochastic cost to the best response dynamic: a revising agent needs to pay switching cost $q$ to switch to a best response and thus does not switch if $q$ exceeds the payoff gain from the switch, i.e., the payoff deficit of the current action. Then, $Q$ is the cumulative distribution function of stochastic switching costs and $[0, \bar{q}]$ is its support.

While we allow asymmetry of the conditional switching rate functions between $Q_{IO}$ and $Q_{OI}$, we often require an exact optimization dynamic to be **monotonic** in the following sense: $Q_{ij}$ does not decrease whenever payoff deficit $\pi_j - \pi_i$ increases and that $Q_{ij}(\boldsymbol{\pi}) > 0$ whenever $\pi_j > \pi_i$. With the above assumption on $Q$, the tempered BRD satisfies this monotonicity.

## 2.3 Generic nonaggregability of heterogeneous evolutionary dynamics

A heterogeneous evolutionary dynamic $\dot{x} = v^F(x)$ is **aggregable** if there is an aggregate dynamic $\bar{v}^F : [0,1] \to [0,1]$ such that

$$\left[ \bar{x}_t = \mathbb{E}_\Theta x_t \text{ and } \dot{x}_t = v^F[x_t] \right] \quad \implies \quad \dot{\bar{x}}_t = \bar{v}^F(\bar{x}_t).$$

That is, evolution of the aggregate strategy $\bar{x}_t$ can be identified from its current state alone—regardless of the underlying strategy composition—and thus it has its own autonomous dynamic $\dot{\bar{x}} = v^F(\bar{x})$.

Aggregability substantially reduces complexity of a heterogeneous dynamic. First, the domain of the aggregate dynamic is just the aggregate strategy space and thus has the finite dimension only of $A - 1$, while that of the heterogeneous dynamic is the space of strategy composition, i.e., a space of joint measures over $\mathcal{A}$ and $\Theta$. Second, the aggregate dynamic indeed allows us to averages off heterogeneity as long as our interest in evolution of the aggregate strategy. Without aggregability, transition of the aggregate strategy depends on the current strategy composition, not only the current aggregate strategy itself.

Ely and Sandholm (2005) prove that the standard BRD is aggregable. However, Zusai (2018a) find that an evolutionary dynamic is not aggregable if difference in the payoff vector among differ-ent types causes difference in their switching rates. This is the case for most of major evolutionary dynamics—such as tempered BRDs, the replicator dynamic, payoff comparison dynamics, since these dynamics have the switching rate continuously changing with the payoff improvement from a switch. The example suggests that this difference may even lead the aggregate strategy from an aggregate equilibrium to another. Below, we quote the generic nonaggregability theorem in Zusai (2018a), applied to our setting.

**Theorem 1** (Generic nonaggregability; Zusai (2018a))**.** *Consider a heterogeneous exact optimization dynamic in a binary-chocie aggregate game with more than one payoff types in $\Theta$. The dynamic is not aggregable and $\dot{\bar{x}}$ is not wholly determined from $\bar{x}$ alone, unless $\bar{x}$ is a corner point either at $0$ or $1$ or the variation in $Q_{IO}(F(\bar{x}), \theta) + Q_{OI}(F(\bar{x}), \theta)$ is zero.*

## 2.4 Stationarity and stability of equilibrium composition

The existence of a unique solution path is proven in Zusai (2018a) for a wide class of evolutionary dynamics in a general aggregate game—including exact optimization dynamics in binary-action aggregate games. The regularity assumptions in Zusai (2018a) reduce to the following two assumptions.

**Assumption 1** (Bounded switching rates)**.** *There exists $\bar{Q} \in \mathbb{R}_+$ such that $Q_{ij}(F(\bar{m}), \theta) \leq \bar{Q}$ for any $m$ in an open interval that contains $[0,1]$ and for any $i, j \in \mathcal{A}$ and $\theta \in \Theta$.*[7]

**Assumption 2** (Lipschitz continuity of the c.d.f. of type)**.** *$P_\Theta$ is Lipschitz continuous.*

He also proves that stationarity of a Nash equilibrium in general under a homogeneous setting is extended to that of a Bayesian equilibrium under a heterogeneous setting. Furthermore, it is proven for a potential game that local asymptotic stability of each individual (isolated) Bayesian equilibrium under "admissible" heterogeneous dynamics is identified from local asymptotic stability of the corresponding aggregate equilibrium under a homogenized BRD.[8] Since a binary-choice action game is a potential game and exact optimization dynamics are admissible, we obtain the following result:

**Theorem 2** (Stationarity and stability of Bayesian equilibrium; Zusai (2018a))**.** *Consider an exact optimization dynamic in a binary-choice game $F$ with the type distribution $\mathbb{P}_\Theta$. Let $\bar{x}^* = \mathbb{E}_\Theta x^*$.*

*i)* *$x^*$ is a stationary state under the exact optimization dynamic if and only if $x^*$ is a Bayesian equilibrium.*

*ii)* *The set of Nash equilibria is globally asymptotically stable.*

*iii)* *$x^*$ is locally asymptotically stable under the exact optimization dynamic if and only if $\bar{x}^*$ is locally asymptotically stable under the homogenized BRD.*

---

[7]For this, the domain of $F$ must be extended to the open interval.

[8]Behind this theorem is the equivalence between a local maximum of the potential function and a local asymptotic stability under admissible dynamics. But we do not use the potential function explicitly for our analysis of a binary game. So we leave it behind Theorem 2.

# 3 Distributional stability and robust critical mass

Theorem 2 suggests that nonaggregability might not alter the set of asymptotically stable equilibria from the one under aggregable dynamics, for example under the standard BRD. Actually, combined with the results of Ely and Sandholm (2005), we can infer that they can be identified just by finding all the asymptotically stable *aggregate* equilibria under the homogeneized BRD.

However, when tracking a dynamic of the aggregate strategy from a given initial state, this does not imply that the long-run outcome from this particular initial state converges to the same stable aggregate equilibrium $x^*$ in nonaggregateble dynamics as in the homogenized BRD. Since the notion of asymptotic stability is defined on the space of strategy composition, a basin of attraction to $x^*$ can be any neighborhood of $x^*$ and thus can require any correlation between participation rates of different types in the initial strategy composition. Suppose that the aggregate strategy is close enough to $\bar{x}^* = \mathbb{E}_\Theta x^*$ so it belongs to the basin of attraction to $\bar{x}^*$ under the homogenized smooth BRD. However, if the strategy composition is largely distorted from the equilibrium composition $x^*$, the strategy composition may not converge to $x^*$ and thus the aggregate strategy may not converge to $\bar{x}^*$; see Example 1. This observation leads to define a stronger concept of local stability.

**Definition 4** (Distributional stability). Aggregate equilibrium $\bar{x}^*$ is **distributionally stable**, if $\mathbb{E}_\Theta x_t$ converges to $\bar{x}^*$ whenever the initial aggregate strategy $\mathbb{E}_\Theta x_0$ is sufficiently close to $\bar{x}^*$, *regardless of the underlying strategy composition $x_0$.*

Note that distributional stability of an aggregate equilibrium generally implies local asymptotic stability of the underlying Bayesian equilibrium; they are equivalent if the dynamic is aggregable or if the Bayesian equilibrium is globally asymptotically stable. Otherwise, distributional stability may be stronger than local stability as shown in the following example. Distributional stability of an aggregate equilibrium requires the basin of attraction to the underlying Bayesian equilibrium to include *all* Bayesian strategies whose aggregate strategies are sufficiently close to the aggregate equilibrium, while local stability of a Bayesian equilibrium allows the basin to exclude unsorted Bayesian strategies that locate far from this Bayesian equilibrium in the space of strategy compositions.

*Example* 1 (Comparison of the BRD and the tBRD in a binary game). We specify the payoff function in a reduced binary game as $F(\bar{x}) = (49\bar{x} - 1)/20$ and the distribution of type $\theta$ by c.d.f. $P_\Theta(\theta) = \sqrt{\theta + 1} - 1$ with support $\Theta = [0,3]$, while assuming $F(\bar{x}) \equiv 0$ at any state $\bar{x}$ and $\theta(\omega) \equiv 0$ for all agents $\omega$.

There are three aggregate equilibria: $\bar{x} = 0, 0.2, 0.25$, as found from Figure 1a. Under the homogenized smooth BRD $\dot{\bar{x}} = P_\Theta(F(\bar{x})) - \bar{x}$, $\bar{x} = 0.2$ is the only unstable equilibria and thus the boundary between the basin of attraction to $\bar{x} = 0$ and that to $\bar{x} = 0.25$.

Let the initial Bayesian strategy $x$ be $x(\theta) = 1$ if $\theta > 33/16$ and $x(\theta) = 0$ for other types. This *reversely* sorted Bayesian strategy yields aggregate equilibrium $\bar{x} = 0.25$. Starting from this Bayesian strategy, the aggregate strategy stays at the stable aggregate equilibrium $\bar{x} = 0.25$ under the heterogeneous standard BRD, which is implied by stationarity of aggregate equilibrium under the homogenized smooth BRD.

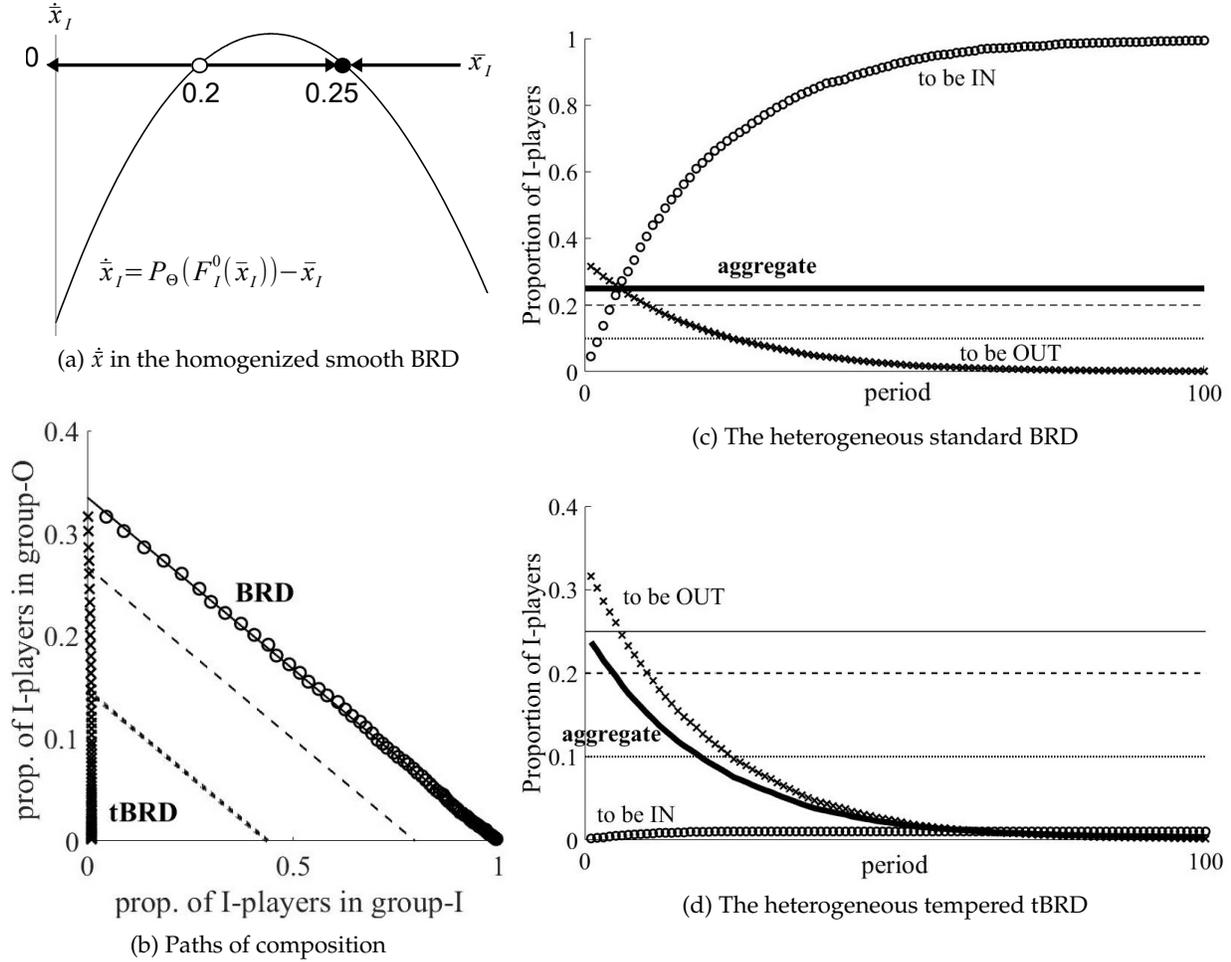

*Figure 1*: The BRD and the tBRD in the binary game in Example 1. In Figure 1b, the thin solid line shows the set of compositions that keep the aggregate strategy to one of aggregate equilibria $\bar{x} = 0.25$; the dashed line corresponds to another aggregate equilibrium $\bar{x} = 0.20$ and the dotted line to $\bar{x} = 0.1$. In Figures 1c and 1d, the horizontal lines show these aggregate equilibria as well.

Under the heterogeneous tBRD, the aggregate mass of action-I players $\bar{x}_t$ leaves $\bar{x} = 0.25$, reaches the robust critical mass $\bar{x} = 0.10$ in a finite time and then stays smaller than this critical mass thereafter. It never returns to $\bar{x} = 0.25$. Therefore, $\bar{x} = 0.25$ is not distributionally stable under the tBRD, while so under the standard BRD. ∎

The logical relationship between distributional stability and local stability suggests to use nonaggregable dynamics to select "robust" equilibria by distributional stability while using the homogenized smooth BRD as the benchmark dynamic at the very first step to narrow down the candidates of robust equilibria.

## 3.1 Distributional critical mass

While evolution of the aggregate strategy generally depends on the underlying strategy composition, there is a possibility that the direction of the aggregate evolution may be unaffected by distributional fluctuation by nonaggregable dynamics at a given aggregate strategy. In the context of entry games, we might expect such a sure movement once the aggregate mass of participants exceed a "critical mass." Below we define the notion of distributional critical masses.

**Definition 5** (Distributional critical mass). $\bar{x}^\dagger$ is a **distributional critical mass** to decrease (resp., increase) $\bar{x}$, if $\mathbb{E}_\Theta \dot{x} = \mathbb{E}_\Theta v^F[x] < 0$ whenever the current aggregate strategy $\mathbb{E}_\Theta x$ coincides with $\bar{x}^\dagger$ regardless of the underlying Bayesian strategy $x$.

In a binary-choice game, distributional stability can be characterized by distributional critical masses.

**Observation 1** (Distributional stability and distributional critical masses). *Let $\bar{x}^\ddagger \in [0,1)$ be a distributional critical mass to increase $\bar{x}$ and $\bar{x}^\dagger \in (0,1]$ be a distributional critical mass to decrease $\bar{x}$. If $\bar{x}^\ddagger < \bar{x}^\dagger$ and there exists only one aggregate equilibrium $\bar{x}^* \in (\bar{x}^\ddagger, \bar{x}^\dagger)$, then $\bar{x}^*$ is distributionally stable.*

The range $[\bar{x}^\ddagger, \bar{x}^\dagger]$ is bound by two distributional critical masses, each of which drives the aggregate strategy toward the inside of this range. Once the aggregate strategy falls into this range, it can no longer escape from this range while the aggregate strategy must converge to some equilibrium according to Theorem 2. Since $\bar{x}^*$ is assumed to be the only one aggregate equilibrium in this range, it must converge to this aggregate equilibrium regardless of the underlying strategy composition. Thus, $\bar{x}^*$ is distributionally stable and the pair of the two critical masses $\bar{x}^\ddagger$ and $\bar{x}^\dagger$ forms a *distributional* basin of attraction $[\bar{x}^\ddagger, \bar{x}^\dagger]$ to the aggregate equilibrium.

## 3.2 How to find a distributional critical mass?

$\bar{x}^\dagger$ is a distributional critical mass to decrease $\bar{x}$, if it satisfies two (jointly sufficient) conditions:

a) $\mathbb{E}_\Theta \dot{x}(\theta) < 0$ at the *perfectly sorted* composition such that
$$x(\theta) = 0 \text{ for any } \theta > P_\Theta^{-1}(\bar{x}^\dagger) \quad \text{and} \quad x(\theta) = 1 \text{ for any } \theta \leq P_\Theta^{-1}(\bar{x}^\dagger).$$

b) This perfectly sorted composition yields the greatest net increase $d\bar{x}/dt$ of action-I players among all strategy compositions such that $\mathbb{E}_\Theta x = \bar{x}^\dagger$

Note that the aggregate strategy in the above perfectly sorted composition is $\mathbb{E}_\Theta x = P_\Theta(P_\Theta^{-1}(\bar{x}^\dagger)) = \bar{x}^\dagger$. We call type $\theta = P_\Theta^{-1}(\bar{x}^\dagger)$ the cut-off type.

This idea leads to the next theorem.

**Theorem 3** (Sufficient condition for a distributional critical mass). *Consider a monotone exact optimization dynamic in a binary aggregate game with positive externality. Assume Assumptions 1 and 2.*

i) Let $\bar{x}^\dagger \in (0,1]$ be an aggregate strategy such that a) $P_\Theta^{-1}(\bar{x}^\dagger) > F(\bar{x}^\dagger)$ and b) $\bar{x}^\dagger = 1$ or $P_\Theta(F(\bar{x}^\dagger)) = 0$ or

$$Q_{IO}(F(\bar{x}^\dagger), P_\Theta^{-1}(\bar{x}^\dagger)) \geq \sup_{\theta \in \Theta_O \cap (-\infty, F(\bar{x}^\dagger))} Q_{OI}(F(\bar{x}^\dagger), \theta). \tag{3}$$

Then, $\bar{x}^\dagger$ is a distributional critical mass to decrease $\bar{x}$.

ii) Let $\bar{x}^\ddagger \in [0,1)$ be an aggregate strategy such that a) $P_\Theta^{-1}(\bar{x}^\ddagger) < F(\bar{x}^\ddagger)$ and b) $\bar{x}^\ddagger = 0$ or $P_\Theta(F(\bar{x}^\ddagger)) = 1$ or

$$Q_{OI}(F(\bar{x}^\ddagger), P_\Theta^{-1}(\bar{x}^\ddagger)) \geq \sup_{\theta \in \Theta_O \cap (F(\bar{x}^\ddagger), +\infty)} Q_{IO}(F(\bar{x}^\ddagger), \theta). \tag{4}$$

Then, $\bar{x}^\dagger$ is a distributional critical mass to increase $\bar{x}$.

Notice that an agent is better off to take O if its type $\theta$, i.e., the outside option from action O is greater than $F(\bar{x}^\dagger)$ and to take I if it is smaller. On the other hand, if the strategy composition is perfectly sorted in the sense that there is a certain cut-off type above which types (outside options) $\theta$ of agents take O and below which take I, then type $\theta^\dagger = P_\Theta^{-1}(\bar{x}^\dagger)$ is the cut-off type when the aggregate mass of $I$ players is $\bar{x}^\dagger$. In part i), the first condition $P_\Theta^{-1}(\bar{x}^\dagger) > F(\bar{x}^\dagger)$ implies that $d\bar{x}/dt < 0$ when the strategy composition is perfectly sorted, since action O is better than action I for the cut-off type. While $x(\theta) = 1$ at all types $\theta \in (F(\bar{x}^\dagger), P_\Theta^{-1}(\bar{x}^\dagger))$ in this perfectly sorted composition, agents of types in this range switch from I to O. There is no other switch in other types, as they have taken the best responses to $\bar{x}^\dagger$: see Figure 2a. Thus, from the perfectly sorted composition, $\bar{x}$ decreases.

Condition (3) guarantees that the perfectly sorted composition yields the greatest net increase $d\bar{x}/dt$ of $I$-players among all strategy compositions that have the aggregate mass of $I$ players equal to $\bar{x}^\dagger$. This condition requires that the switching rate of the cut-off type is greater than the switching rate of any type who switches to I when the payoff of action $I$ is $F(\bar{x}^\dagger)$. Any unsorted composition has a positive mass of action-I players whose types are above the cut-off type and also a positive mass of action-O players whose types are below the cut-off type: see Figure 2b. If (3) is satisfied, the former group of agents switches from I to O at greater switching rates than the latter group switches from O to I. Thus, the deviation from perfect sorting lowers $d\bar{x}/dt$ further from $d\bar{x}/dt < 0$ at the perfectly sorted composition.

The first condition a) $P_\Theta^{-1}(\bar{x}^\dagger) > F(\bar{x}^\dagger)$ in part i) is indeed a sufficient condition for $d\bar{x}/dt < 0$ under the homogenized smooth BRD. Because a distributional critical mass requires the additional condition b), robustness to the underlying strategy composition in nonaggregable dynamics requires a stronger condition for $d\bar{x}/dt < 0$ than in the aggregable dynamic.

Note that condition (3) is not imposed in the following two cases. First, if the aggregate strategy is at a corner, i.e., $\bar{x}^\dagger$ is exactly 0 or 1, then the perfectly sorted composition is the only composition to have the designated aggregate strategy and thus the additional condition (3) is trivially not needed. Second, if the type distribution $P_\Theta$ assigns (almost) no agent to types which would switch actions oppositely to the cut-off type, then (almost) all agents have the same best response as the cut-off type; the direction of change in the aggregate strategy is trivially determined by looking at the best response of the cut-off type and thus by comparing $\theta = P_\Theta^{-1}(\bar{x}^\dagger)$ and $F(\bar{x}^\dagger)$. This is the case when $P_\Theta(F(\bar{x}^\dagger)) = 0$.

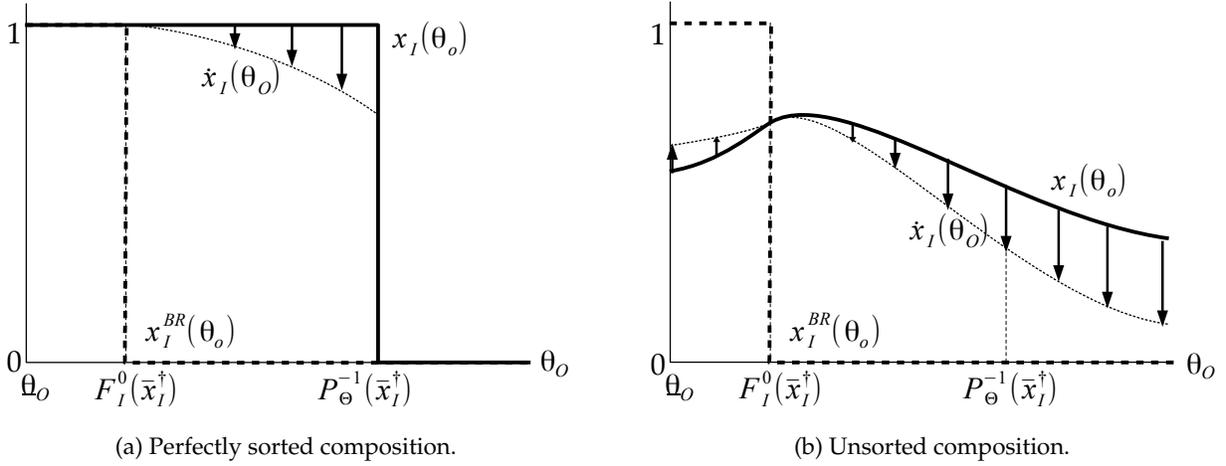

(a) Perfectly sorted composition.   (b) Unsorted composition.

*Figure 2*: Transition of Bayesian strategy under $P_\Theta^{-1}(\bar{x}^\dagger) > F(\bar{x}^\dagger)$. Recall that $P_\Theta^{-1}(\bar{x}^\dagger)$ is the cut-off type. In each figure, the bold solid lines show the initial Bayesian strategy $x$. Arrows show the transition of Bayesian strategy $\dot{x}$ at each type. The bold dashed line shows $x^{BR}$, the Bayesian strategy in which every agent takes the best response for its own type to $\bar{x}^\dagger$.

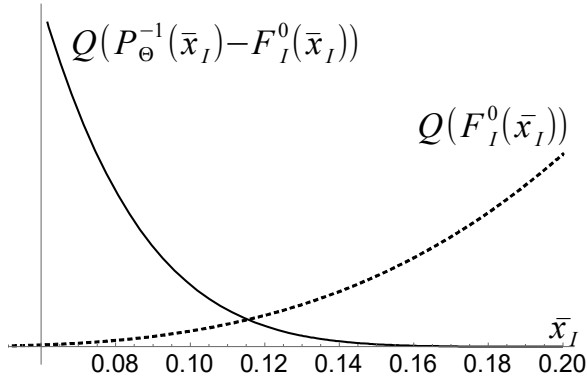

*Figure 3:* Condition (3) to find the robust critical mass

*Example* 2 (Distributional critical mass in Example 1.). Revisit the tBRD with $Q(\breve{\pi}) = \breve{\pi}^3$ in Example 1. Figure 3 shows the graphs of the LHS and the RHS of (3), as the support of $\theta$ is $[0,3]$ and thus the supremum in the RHS is achieved at $\theta = 0$. From this figure, we can see that aggregate strategy $\bar{x} = 0.10$ satisfies condition (3) for a distributional critical mass to decrease $\bar{x}$. Thus $\bar{x} = 0$ is distributionally stable, with a distributional basin of attraction $[0, 0.10]$. ∎

### 3.3 A few caveats

First, even if there is only one distributionally stable aggregate equilibrium and the set of aggregate equilibria is globally asymptotically stable, it does not imply that this aggregate equilibrium is globally stable; this is unlike the relationship between a unique local stable state and global stability. In the last example, the Bayesian equilibrium $x^*$ that underlies $\bar{x}^* = 0.25$ is locally stable; thus, if the initial *Bayesian* strategy is sufficiently close to $x^*$, then it converges to $x^*$. Hence, the corresponding aggregate strategy converges to $\bar{x}^*$ as well, not to the unique distributionally stable aggregate equilibrium.

Second, even if a pair of distributional critical masses forms a basin of attraction to a distributionally stable equilibrium $\bar{x}^*$ and the dynamic starts from this basin, $\bar{x}$ may not monotonically converge to $\bar{x}^*$. The conditions (3) and (4) may not be satisfied near an aggregate equilibrium and thus the direction of change in the aggregate strategy may become dependent on the underlying strategy composition when it travels far from distributional critical masses and comes close to the aggregate equilibrium. For Example 1, check from Figure 6 that distributional critical masses do not exist near aggregate equilibrium $x = 0.25$ since the payoff deficit $|F(\bar{x}) - P_\Theta^{-1}(\bar{x})|$ of the cut-off type becomes small there, unless $\check{\pi}^\sharp$ is exactly zero.

Actually, the following theorem shows that fluctuation of the aggregate strategy persists when the Bayesian strategy comes close to a Bayesian equilibrium.

**Theorem 4** (Fluctuation around aggregate equilibrium). *Consider a tBRD in a (reduced) binary game. Suppose that $P_\Theta$ is a continuous distribution with density $p_\Theta(\theta) > 0$ at all $\theta \in \Theta_O$.*

*Consider a Bayesian strategy $x^*$; let $X^* = \int x^* d\mathbb{P}_\Theta$ be the corresponding strategy composition. Suppose that $x^*$ satisfies either one of the following two conditions. Then, for any small distance from $x^*$, there is a direction of change from $x^*$ that results in further deviation of the aggregate strategy from $\bar{x}^* := \mathbb{E}_\Theta x^*$: for any $\bar{\varepsilon} > 0$, there is a Bayesian strategy $x^\dagger$ with strategy composition $X^\dagger = \int x^\dagger d\mathbb{P}_\Theta$ such that $d(X^\dagger, X^*) < \bar{\varepsilon}$ in Prokhorov metric on $\mathcal{X}$, $\bar{x}^\dagger := \mathbb{E}_\Theta x^\dagger > \bar{x}^*$ and $\dot{\bar{x}} = \mathbb{E}_\Theta v^F[\bar{x}^\dagger] > 0$.*[9]

1. *$x^*$ is not a Bayesian equilibrium but its aggregate $\bar{x}^*$ is an aggregate equilibrium; besides, $x^*$ satisfies the balancing condition (7). Further, $\frac{dF}{d\bar{x}}(\bar{x}^*) > 0$.*

2. *$x^*$ is a Bayesian equilibrium, with $F(\bar{x}^*)$ belonging to the interior of $\Theta_O$.*

In the aggregate BRD, stability of an aggregate equilibrium is characterized by $dF/d\bar{x}$ and $dP_\Theta^{-1}/d\bar{x} = 1/p_\Theta(P_\Theta^{-1}(\bar{x}))$. In particular, if $dF/d\bar{x}$ is greater than $dP_\Theta^{-1}/d\bar{x} > 0$ at an aggregate equilibrium, this aggregate equilibrium is unstable (Sandholm, 2007, Theorem 4.1). This holds in the aggregation of the tempered BRD as well, according to part 1 of the above theorem.

Somewhat surprisingly, an *aggregate* strategy generally oscillates according to the second part, however close the underlying Bayesian strategy is to a Bayesian equilibrium and whatever $dF/d\bar{x}$ is; in any neighborhood of a Bayesian equilibrium $x^*$, we can find a Bayesian strategy at which the aggregate strategy move away temporarily from $\bar{x}^* = \mathbb{E}_\Theta x^*$. However, this theorem does not mean instability of *Bayesian equilibrium*. Even when the Bayesian strategy converges to a Bayesian equilibrium, aggregate strategy may fluctuate due to sorting of an unbalanced strategy composition.

---

[9]Thanks to Lipschitz continuity of $v^F$, a Bayesian strategy $x$ that is sufficiently close to $x^\dagger$ also satisfies these properties: $\bar{x} = \mathbb{E}_\Theta x > \bar{x}^*$ and $\dot{\bar{x}} = \mathbb{E}_\Theta v^F(x) > 0$.

# 4 Distributional instability

## 4.1 Switching rate distribution as a summary measure

Nonaggregability implies that the underlying strategy composition matters for evolution of the aggregate strategy. In a binary-choice game, the former has as a high dimension as the type space while the latter reduces to a single dimension. As we are interested in this lower dimensional dynamic of the aggregate strategy, it is natural to search for a "summary statistics" that captures crucial factors of the strategy composition well enough to identify the dynamic of the aggregate strategy. The generic nonaggregability theorem (Theorem 1) suggests that switching rates of different types of agents are crucial for nonaggregability. Developing this idea, we focus on the distribution of switching rates over heterogeneous agents; we will find that it fully pins down the dynamic of the aggregate strategies. The switching rate distribution also provides simple sufficient conditions for both the medium-run and long-run escape from an aggregate equilibrium and thus help us to see what happens if an aggregate equilibrium is not distributionally stable.

Since Theorems 5 and 6 below hold for aggregate games with arbitrarily finitely many actions, let us introduce the notation for the general aggregate game. Denote the finite action set by $\mathcal{A} = \{1, \ldots, A\}$, the type space by $\Theta$, and the payoff for type $\theta$ from action $a \in \mathcal{A}$ given aggregate strategy $\bar{x} \in \Delta^A := \{\bar{x} \in \mathbb{R}^A \mid \sum_{a \in \mathcal{A}} \bar{x}_a = 1\}$ by $F_a[\bar{x}](\theta)$. In the binary aggregate game, $F_I[\bar{x}](\theta) = F_I(\bar{x})$ and $F_O[\bar{x}](\theta) = \theta$.

For each action $a \in \mathcal{A}$, we look into the mass of agents who take the action $a$ when the aggregate strategy arrives exactly at aggregate equilibrium $\bar{x}^*$, say time 0. Let $\mathfrak{a}_0 : \Omega \to \mathcal{A}$ be the action profile at this moment of time. Let $\beta_b^{-1}(\bar{x}^*)$ be the set of types for which action $b$ is the unique best response given the aggregate state $\bar{x}^*$: in our binary aggregate game, $\beta_O^{-1} = (F(\bar{x}^*), +\infty)$ and $\beta_I^{-1} = (-\infty, F(\bar{x}^*))$. We identify the agents for who currently take action $a$ but have the *unique best response* to $\bar{x}^*$ *other than a*:

$$\Omega_{a,0}^O := \bigcup_{j \in \mathcal{A} \setminus \{a\}} \{\omega \in \Omega : \theta(\omega) \in \beta_j^{-1}(\bar{x}^*) \text{ and } \mathfrak{a}_0(\omega) = a\}.$$

Call this set of agents $\Omega_{a,0}^O$ the **source of outflows** from $a$, as those agents will leave the mass of action-$a$ players sooner or later, as long as the aggregate strategy remains at $\bar{x}^*$ and the payoff vector does not change. Similarly, we identify agents for who have action $a$ as the unique best response to $\bar{x}^*$ but *are not taking* it at time 0:

$$\Omega_{a,0}^I := \bigcup_{i \in \mathcal{A} \setminus \{a\}} \{\omega \in \Omega : \theta(\omega) \in \beta_a^{-1}(\bar{x}^*) \text{ and } \mathfrak{a}_0(\omega) = i\}.$$

Call this $\Omega_{a,0}^I$ the **source of inflows** to $a$.

Aggregate equilibrium implies that the aggregate masses of the source of outflows and of the source of inflows are equal to each other:

$$\mathbb{P}_\Omega(\Omega_{a,0}^O) = \mathbb{P}_\Omega(\Omega_{a,0}^I) \qquad \text{for each } a \in \mathcal{A}.$$

The next theorem suggests that, unless switching rate $Q_{\cdot\cdot}$ is constant to payoff, stationarity of aggregate equilibrium requires more than the balance in these aggregate masses. To precisely state the stationarity condition, we define the **distributions of switching rates** $Q_{..}(F(\bar{x}^*), \theta)$ in the source of inflows $\check{Q}^I_{a,0}$ and in the source of outflows $\check{Q}^O_{a,0}$ as follows:

$$\check{Q}^I_{a,0}(q) := \sum_{i \in \mathcal{A} \setminus \{a\}} \mathbb{E}_\Theta \left[ \mathbf{1}\left(\theta \in \beta_a^{-1}(\bar{x}^*)\right) \mathbf{1}\left(Q_{ia}(F(\bar{x}^*), \theta) \leq q\right) x_0(\theta) \right], \tag{5a}$$

$$\check{Q}^O_{a,0}(q) := \sum_{j \in \mathcal{A} \setminus \{a\}} \mathbb{E}_\Theta \left[ \mathbf{1}\left(\theta \in \beta_j^{-1}(\bar{x}^*)\right) \mathbf{1}\left(Q_{aj}(F(\bar{x}^*), \theta) \leq q\right) x_{a,0}(\theta) \right] \tag{5b}$$

for each $q \in \bar{\mathbb{R}} := \mathbb{R} \cup \{-\infty, +\infty\}$. For example, in $\check{Q}^I_{a,0}(q)$, we first restrict attention to the types who have $a$ as the unique best response and count the mass of these types of agents who are taking other actions $i \in \mathcal{A} \setminus \{a\}$ as long as their switching rates from current action $i$ to the best response $a$ do not exceed $q$. By changing $q$ and then collecting $\check{Q}^I_{a,0}(q)$ over all $q \in \bar{\mathbb{R}}$, we can construct the cumulative distribution function of the switching rates $Q_{..}(F(\bar{x}^*), \theta)$ in the source of inflows, i.e., $\check{Q}^I_{a,0} : \bar{\mathbb{R}} \to \mathbb{R}_+$. Similarly, we construct the c.d.f. of the switching rates in the source of outflows, $\check{Q}^O_{a,0} : \bar{\mathbb{R}} \to \mathbb{R}_+$. Note $\check{Q}^I_{a,0}(+\infty) = \mathbb{P}_\Omega(\Omega^I_{a,0})$ and $\check{Q}^O_{a,0}(+\infty) = \mathbb{P}_\Omega(\Omega^O_{a,0})$.

The transition of $\bar{x}_a$ is obtained as the difference in the aggregate of switching rates in the source of inflows and the aggregate of that in the source of outflows.

**Theorem 5** (Aggregate transition identified from switching rate distributions). *The dynamic of the aggregate strategy is fully identified from $\check{Q}^I$ and $\check{Q}^O$*

$$\dot{\bar{x}}_{a,0} = \int_{-\infty}^{+\infty} q \check{Q}^I_{a,0}(dq) - \int_{-\infty}^{+\infty} q \check{Q}^O_{a,0}(dq). \tag{6}$$

## 4.2 Non-stationarity of aggregate equilibrium and short-run escape from an aggregate equilibrium

The switching rate distributions in the sources of inflows and outflows provide a simple and intuitive condition for stationarity of an aggregate equilibrium; further, when the aggregate strategy leaves an aggregate equilibrium, this notion further tells whether the aggregate strategy returns to this aggregate equilibrium in a finite time.

First of all, (6) tells that $\dot{\bar{x}}_{a,0} = 0$ if the switching rate distribution in the sources of inflows coincides that in the sources of outflows, i.e., $\check{Q}^O_{a,0} \equiv \check{Q}^O_{a,0}$. The next theorem indeed confirms that coincidence of these distributions of switching rates is not only sufficient but *necessary* for stationarity of the aggregate strategy. The condition for aggregate equilibrium can be rephrased as

$$\check{Q}^I_{a,0}(+\infty) = \check{Q}^O_{a,0}(+\infty) \qquad \text{for each } a \in \mathcal{A}.$$

We can see immediately by comparing this equation with the identity (7) that an aggregate equilibrium is not guaranteed to be stationary.

**Theorem 6** (Stationarity of aggregate equilibrium in exact optimization dynamics). *Consider an exact optimization dynamic. Assume Assumptions 1 and 2. Let $\bar{x}^* \in \Delta^A$ be an aggregate equilibrium. The* *trajectory of the aggregate strategy $\{\bar{x}_t\}$ stays at $\bar{x}^*$ for time interval $[0, \Delta T]$ (with any $\Delta T > 0$), if and only if the underlying Bayesian strategy $x_0$ at time $0$ satisfies the detailed balancing condition:*

$$\check{Q}^I_{a,0}(q) \equiv \check{Q}^O_{a,0}(q) \qquad \text{for all } q \in \bar{\mathbb{R}}, a \in \mathcal{A}. \tag{7}$$

In the standard BRD, the condition reduces to the aggregate equilibrium condition, as was verified by Ely and Sandholm (2005). If the tempering function $Q$ of a tBRD is strictly increasing over all $\mathbb{R}_+$, it reduces to the coincidence between the distributions of *payoff deficits* in the source of inflows and in the source of outflows, defined by

$$\check{\Pi}^I_{a,0}(\check{\pi}) = \sum_{i \in \mathcal{A} \setminus \{a\}} \mathbb{E}_\Theta \left[ \mathbf{1}\left(\theta \in \beta_a^{-1}(\bar{x}^*)\right) \mathbf{1}\left(\check{F}_i[\bar{x}^*](\theta) \leq \check{\pi}\right) x_0(\theta) \right]$$

$$\check{\Pi}^O_{a,0}(\check{\pi}) = \mathbb{E}_\Theta \left[ \mathbf{1}\left(\theta \notin \beta_a^{-1}(\bar{x}^*)\right) \mathbf{1}\left(\check{F}_a[\bar{x}^*](\theta) \leq \check{\pi}\right) x_{a,0}(\theta) \right]$$

for each $a \in \mathcal{A}$ and $\check{\pi} \in \bar{\mathbb{R}}$. Notice that this balancing condition for the tBRD is irrelevant of specification of the tempering function $Q$, except the assumption of strict increasing.

**Corollary 1.** *Assume the assumptions in Theorem 6.*

1. *In the standard BRD, (7) reduces to $\mathbb{P}_\Omega(\Omega^I_{a,0}) = \mathbb{P}_\Omega(\Omega^O_{a,0})$ for each $a \in \mathcal{A}$. So, once aggregate strategy reaches any aggregate equilibrium, it never leaves the aggregate equilibrium whatever the underlying Bayesian strategy is.*

2. *In a tBRD with tempering function $Q$ strictly increasing over all $\mathbb{R}_+$, (7) reduces to $\check{\Pi}^I_{a,0} \equiv \check{\Pi}^O_{a,0}$ for each $a \in \mathcal{A}$.[10]*

Technically, the key idea behind Theorem 6 is that the aggregate inflow and outflow of each action's players at each time $t \in [0, \Delta T]$ can be seen as the moment generating functions of the switching rates distributions in the sources of outflows and of inflows. We can roughly say that, at each level of the switching rate and in each of the aggregate inflow and outflow, the proportion of the agents whose switching rates match with this level in the aggregate flow changes over time. If the switching rate is high, these agents switch the action early; in the aggregate flow, such agents are more dominant early but become extinct later. The change of the aggregate flow over time reflects these changes of these proportions of different types in the aggregate flow; mathematically, the aggregate flow as a function of time can be interpreted as the moment generating function of the switching rate distribution in the source of this aggregate flow. Even if we look only at a short time interval, it generates enough variation to identify the distribution.

Imagine an economist who is interested in the strategy composition over heterogeneous agents, though the economist can observe only the aggregate strategy. Theorem 6 suggests that, if the aggregate strategy does not show fluctuation any more, the composition indeed is balanced though it may not be completely sorted. If the aggregate strategy oscillates, the oscillation implies that the adjustment of the underlying composition is still under way toward a Bayesian equilibrium.

---

[10]For a binary aggregate game, the condition $\check{\Pi}^I_{I,0} \equiv \check{\Pi}^O_{I,0}$ is equivalent to $\check{\Pi}^I_{O,0} \equiv \check{\Pi}^O_{O,0}$ and they reduce to $\int_{F(\bar{x}^*)-\check{\pi}}^{F(\bar{x}^*)} x_{O,0}(\theta) dP_\Theta(\theta) \equiv \int_{F(\bar{x}^*)}^{F(\bar{x}^*)+\check{\pi}} x_{I,0}(\theta) dP_\Theta(\theta)$ for all $\check{\pi}$; denoting the density function of $P_\Theta$ by $p_\Theta$, this further reduces to $x_{O,0}(F(\bar{x}^*) - \check{\pi}) p_\Theta(F(\bar{x}^*) - \check{\pi})) = x_{I,0}(F(\bar{x}^*) + \check{\pi}) p_\Theta(F(\bar{x}^*) + \check{\pi}))$.

## 4.3 Stochastic dominance and escape from aggregate equilibrium

What happens if the detailed balancing condition (7) does not hold when the aggregate strategy arrives at an aggregate equilibrium? We can still predict the transition of the aggregate strategy $\dot{\bar{x}}_0$ just at this moment of time using equation (6). If the aggregate switching rate in the source of inflows to action-$a$ players exceeds that in the source of outflows, then the aggregate mass of action-$a$ players should increase at least at this moment of time. If $F_a(\bar{x})$ increases with $\bar{x}_a$ due to positive externality, then this deviation from the aggregate strategy is reinforced; however, it is possible that this will be reversed later by agents with smaller switching rates if the source of inflows have a smaller mass of such agents than the source of outflows; these agents will become more dominant in the transition of the aggregate strategy as time passes.

Stochastic ordering between the switching rate distributions in the source of inflows and in the source of outflows prohibits such reverse, whichever dominates the other. In the next theorem, we confirm this intuition and verify permanent deviation from an aggregate equilibrium (in any long but finite range of time) by focusing on a binary game. Although it does not negate the possibility of returning to the initial aggregate equilibrium *in the limit state*, the theorem tells that it is only asymptotic.

**Theorem 7** (Escape from aggregate equilibrium with no return in a finite time). *Consider a motonone exact optimization dynamic in a (reduced) binary game with positive externality. Assume Assumptions 1 and 2.*

*Let the aggregate strategy be exactly at an aggregate equilibrium $\bar{x}^* \in \Delta^A$ at time 0, i.e., $\bar{x}_0 = \bar{x}^*$. Suppose that the switching rate distribution in the source of outflows of action-I players $\check{Q}^O_{I,0}$ dominates [is dominated by, resp.] that in the source of inflows of action-I players $\check{Q}^I_{I,0}$ at time 0 in the sense of second-order stochastic domination. Then $\bar{x}_t$ is always smaller [greater, resp.] than $\bar{x}^*$ at any $t \in (0, \infty)$.*[11]

## 4.4 Long-run escape from "stable" aggregate equilibrium

If $\check{Q}^O_{I,0}$ stochastically dominates $\check{Q}^I_{I,0}$, then the aggregate strategy escapes from $\bar{x}_0 = \bar{x}^*$ and indeed stays lower than it. In the next corollary, condition (9) assures that such an escaping path of the aggregate strategy $\bar{x}$ reaches the distributional critical mass $\bar{x}^\dagger$ in a finite time and thus it never returns to $\bar{x}^*$ even asymptotically.

**Theorem 8** (Long-run escape from an aggregate equilibrium). *Consider the situation in part i) of Theorem 3; $\bar{x}^\dagger$ is a distributional critical mass to decrease $\bar{x}$. Let the initial aggregate strategy at time 0 be exactly at aggregate equilibrium $\bar{x}^* \in [\bar{x}^\dagger, 1]$. Denote by $\check{Q}^I_{I,0}$ and $\check{Q}^O_{I,0}$ the switching rate distributions in the source of inflows to action I and in the source of outflows from I. Define $\bar{\bar{x}}_t \in [0,1]$ by*

$$\bar{\bar{x}}_t := \bar{x}^* - \int_0^t \int_0^\infty e^{-q\tau} \check{Q}^I_{I,0}(dq) d\tau + \int_0^t \int_0^\infty e^{-qt} \check{Q}^O_{I,0}(dq) d\tau \tag{8}$$

---

[11]Distribution $\check{Q}^O_{I,0}$ on $\mathbb{R}$ dominates $\check{Q}^I_{I,0}$ in the sense of second-order stochastic domination if $\int_{-\infty}^{\bar{q}} \check{Q}^O_{I,0}(q)dq \geq \int_{-\infty}^{\bar{q}} \check{Q}^I_{I,0}(q)dq$ for any $\bar{q} \in \mathbb{R}$ with strict inequality over some interval. This is weaker than first-order stochastic dominance, i.e., $Q^O_{I,0}(\bar{q}) \geq \check{Q}^I_{I,0}(\bar{q})$ for any $\bar{q} \in \mathbb{R}$ with strict inequality over some interval.

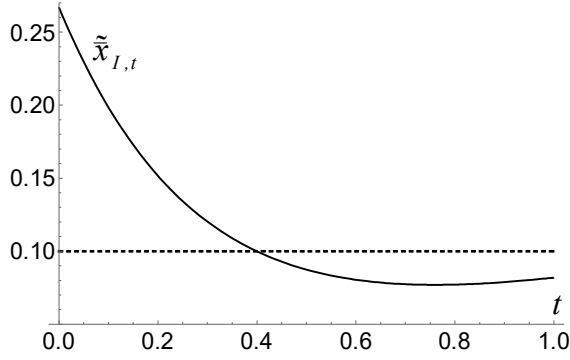

*Figure 4:* Upper bound on $\bar{\tilde{x}}_t$

for each $t \in \mathbb{R}_+$.

If there exists $t \in \mathbb{R}_+$ such that

$$\bar{\tilde{x}}_\tau < \bar{x}^* \quad \text{for any } \tau \in (0, t] \quad \text{and} \quad \bar{\tilde{x}}_t < \bar{x}^\dagger, \tag{9}$$

then the aggregate mass of action-I players $\bar{x}$ first gets smaller than $\bar{x}^*$, reaches the distributional critical mass $\bar{x}^\dagger$ in a finite time, and stays smaller than $\bar{x}^\dagger$ forever since then.

Given $\bar{x}_t < \bar{x}_0 = \bar{x}^*$, positive externality implies that the incentive to switch to $I$ becomes smaller at time $t$ than at time 0, while the incentive to switch to $O$ becomes greater. Thus, the switching rates for those who switch to $I$ at time $t$ should be smaller than the rates at time 0, which are collected in $\check{Q}^I_{I,0}$; the switching rates become greater than at time 0 for those who switch to $O$. In (8), $\bar{\tilde{x}}_t$ represents the aggregate strategy at time $t$ in a hypothetical situation where every agent's switching rate was unchanged from time 0. It gives an upper bound on the actual aggregate strategy $\bar{x}_t$, because these changes in the switching rates induce more switches to $O$ and less switches to $I$ in the actual transition of $\bar{x}_t$ than in this hypothetical transition of $\bar{\tilde{x}}_t$. Thus, if even this upper bound $\bar{\tilde{x}}_t$ hits the distributional critical mass, so does the actual aggregate strategy $\bar{x}_t$.

According to Theorem 7, second-order stochastic domination of $\check{Q}^O_{I,0}$ over $\check{Q}^I_{I,0}$ is sufficient to meet the first condition $\bar{\tilde{x}}_\tau < \bar{x}^*$ for all $\tau \in (0, +\infty)$, though not necessary to meet it for a finite interval of time $(0, t]$ as in (9). However, here the additional condition $\bar{\tilde{x}}_t < \bar{x}^\dagger$ requires the escape $\bar{x}^* - \bar{\tilde{x}}_t$ from the initial aggregate equilibrium is large enough to hit the distributional critical mass.

*Example* 3 (Condition for long-run escape in Example 1). The Bayesian strategy $x_0(\theta) = \mathbf{1}\{\theta > 33/16\}$ satisfies condition (9) for the heterogeneous tBRD, as confirmed below, and thus the tBRD from this Bayesian strategy arrives the distributional critical mass $\bar{x}^\dagger = 0.10$ in a finite time and never returns to $\bar{x}^* = \mathbb{E}_\Theta x_0 = 0.25$

Under Bayesian strategy $x(\theta) = \mathbf{1}(\theta > 33/16)$, the aggregate strategy is $\bar{x} = 1/4$ and thus $F(\bar{x}) = P^{-1}_{\Theta_O}(\bar{x}) = 9/16$. Hence, the source of inflows to action-$I$ players consists of all the agents with $\theta < 9/16$, who all initially take action $O$. The source of outflows from action $I$ consists of all the agents with $\theta > 33/16$, who all initially take action $I$. The distributions of payoff deficits $\check{\Pi}^I_{I,0}$ and $\check{\Pi}^O_{I,0}$ have densities such as $\check{\pi}^I_{I,0}(\check{\pi}) = p_\Theta(9/16 - \check{\pi})$ for all $\check{\pi} = 9/16 - \theta \in [9/16 -$

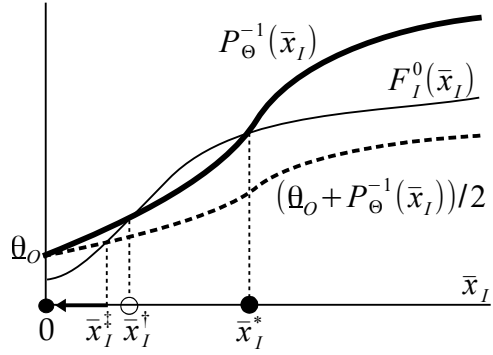

*Figure 5*: The common payoff function and the inverse c.d.f. of the type distribution in a binary coordination ASAG.

$9/16, 9/16 - 0] = [0, 9/16]$ and $\check{\pi}_{I,0}^O(\check{\pi}) = p_\Theta(9/16 + \check{\pi})$ for all $\check{\pi} = \theta - 9/16\pi \in [33/16 - 9/16, 3 - 9/16] = [3/2, 39/16]$. Therefore, we have

$$\bar{x}_t = \bar{x}^* - \int_0^t \left\{ \int_0^{9/16} \exp(-\check{\pi}^3 \tau) \frac{d\check{\pi}}{\sqrt{9/16 - \check{\pi}}} - \int_{3/2}^{39/16} \exp(-\check{\pi}^3 \tau) \frac{d\check{\pi}}{\sqrt{9/16 + \check{\pi}}} \right\} d\tau$$

$$\leq \bar{x}^* - \int_0^t \left\{ e^{-(9/16)^3 \tau} \int_0^{9/16} \frac{d\check{\pi}}{\sqrt{9/16 - \check{\pi}}} - e^{-(3/2)^3 \tau} \int_{3/2}^{39/16} \frac{d\check{\pi}}{\sqrt{9/16 + \check{\pi}}} \right\} d\tau.$$

Figure 4 shows the graph of this upper bound on $\bar{x}_t$, denoted by $\tilde{\bar{x}}_t$. From this graph, we can see that $\tilde{\bar{x}}_t$ goes lower than $\bar{x}^\dagger = 0.1$ when $t$ is around 0.5 (and greater); so does $\bar{x}_t$, possibly earlier. Therefore, condition (9) is satisfied at this $t$. ∎

One might wonder a more direct condition for the long-run escape. Below we focus on a binary ASAG with two interior aggregate equilibria. We set the initial strategy composition to a *reversed* strategy composition in which those who have greater values of $\theta_O$ are initially chosing $I$, while setting its aggregate strategy to the interior stable aggregate equilibrium $\bar{x}^*$ under the homogenized smooth BRD. In this extreme case, we find that, if the switching rate function is sensitive enough to payoff gains, then the aggregate strategy escapes from $\bar{x}^*$ and converges to a corner equilibrium $\bar{x} = 0$.

Specifically, consider a binary ASAG such as

$$F_I^0(\bar{x}_I) \begin{cases} > \\ = P_\Theta^{-1}(\bar{x}_I) \\ < \end{cases} \begin{array}{l} \text{if } \bar{x}_I \in (\bar{x}_I^\dagger, \bar{x}_I^*), \\ \text{if } \bar{x}_I \in \{\bar{x}_I^\dagger, \bar{x}_I^*\}, \\ \text{if } \bar{x}_I \notin [\bar{x}_I^\dagger, \bar{x}_I^*], \end{array}$$

where $0 < \bar{x}_I^\dagger < \bar{x}_I^* < 1$. Assume strict increasingness of $F_I^0$ (positive externality of action $I$) and continuous type distribution of $\theta_O$, i.e., continuity of c.d.f. $P_\Theta$, as well as Lipschitz continuity of $F_I^0$. Under the homogenized smooth BRD, $\bar{x}_I = 0$ and $\bar{x}_I = \bar{x}_I^*$ are stable aggregate equilibria and $\bar{x}_I = \bar{x}_I^\dagger$ is an unstable one. Further, we assume that there is a lower bound $\underline{\theta}_O$ on the type space $\Theta_O \subset \mathbb{R}$. We call this game a **binary coordination ASAG**.

The initial Bayesian strategy $\mathbf{x}^0$ is set as

$$x_I^0(\theta_O) = \begin{cases} 1 & \text{if } \theta_O > P_\Theta^{-1}(1 - \bar{x}_I^*), \\ 0 & \text{if } \theta_O < P_\Theta^{-1}(1 - \bar{x}_I^*). \end{cases}$$

In this composition, those who have *relatively high* values of the outside option $\theta_O$ happen to choose $I$ while those who have lower values of the outside option happen to choose $O$; so their choices are initially opposite to their current best responses. We call this composition a *reversed composition*.[12] The type $\hat{\theta}_O^0 := P_\Theta^{-1}(1 - \bar{x}_I^*)$ is the threshold between initial action-$I$ players and $O$ players. Assume $\hat{\theta}_O^0 > \theta_O^* := F_I^0(\bar{x}_I^*)$.

Yet, the aggregate strategy in this strategy composition coincides with aggregate equilibrium $\bar{x}_I^*$. Thus, the aggregate strategy must stay there under the homogenized smooth BRD; since it is also a stable equilibrium, it cannot leave this aggregate equilibrium even if there is so a small perturbation that moves the aggregate strategy $\bar{x}_I$ above $\bar{x}_I^\dagger$. Since the heterogeneous standard BRD aggregates to the homogeneous smooth BRD, the underlying strategy composition converges to the corresponding Bayesian equilibrium $\mathbf{x}^*$ such that $x_I^*(\theta_O) = 1$ for any $\theta_O < \theta_O^*$ and 0 for any $\theta_O > \theta_O^*$.

However, the next theorem suggests that, under a nonaggregable dynamic, the aggregate strategy may escape from the "stable" aggregate equilibrium $\bar{x}_I^*$ and it may even converge to another aggregate equilibrium $\bar{x}_I = 0$. This depends on the difference in switching rates between those who switch from I to O and those who switch from O to I. The next theorem presents a rough sufficient condition that allows us to predict the escape just by comparing the switching rate of the threshold type $\hat{\theta}_O^0$ and that of the lowest type $\theta_O = \underline{\theta}_O$ at time 0.

**Corollary 2** (Escape from a "stable" aggregate equilibrium in a binary coordination ASAG). *Consider a monotonic exact optimization dynamic in a binary coordination ASAG, starting from the reversed composition $x^0$ at time 0. Assume*

$$\hat{\theta}^0 := P_\Theta^{-1}(1 - \bar{x}^*) > \theta^* := F^0(\bar{x}^*)$$

*and*

$$r := Q_{OI}(F^0(\bar{x}^*), \underline{\theta})/Q_{IO}(F^0(\bar{x}^*), \hat{\theta}^0) < 1.$$

*Then, the following holds.*

i) *$\bar{x}$ decreases from $\bar{x}^*$ at least temporarily: $d\bar{x}/dt < 0$ at time 0.*

ii) *Let $\bar{x}^\ddagger$ be an aggregate strategy such that*

$$Q_{IO}(F^0(\bar{x}), P_\Theta^{-1}(\bar{x})) \geq Q_{OI}(F^0(\bar{x}), \underline{\theta}) \qquad \text{whenever } \bar{x} \leq \bar{x}^\ddagger. \tag{10}$$

---

[12]This reversed composition is obviously the most extreme case that is farthest from the equilibrium composition. However, the Lipschitz continuity of $\mathbf{v}^F$ in **??** implies robustness of our claims in Corollary 2. Although it does not directly guarantee the robustness of a long-run outcome, it implies that the aggregate strategy reaches the robust critical mass $\bar{x}_I^\ddagger$ in a finite time as long as the initial strategy composition is sufficiently close to $\mathbf{x}^0$. Since then, it converges to $\bar{x}_I = 0$.

*If the ratio of the initial switching rates r satisfies*

$$\bar{x}_I^*(1 - (1-r)r^{r/(1-r)}) \leq \bar{x}^{\ddagger}, \tag{11}$$

*then the solution trajectory from the reversed composition* $\mathbf{x}^0$ *converges to* $\bar{x} = 0$.

These conditions are satisfied in our canonical example, Example 1. Condition (10) implies that $\bar{x}_I^{\ddagger}$ is a robust critical mass to decrease $\bar{x}_I$. Note that, under a tempered BRD, the condition (10) for $\bar{x}_I^{\ddagger}$ reduces to

$$F_I^0(\bar{x}_I) \leq 0.5 \left( P_{\Theta}^{-1}(\bar{x}_I) + \underline{\theta}_O \right) \qquad \text{whenever } \bar{x}_I \leq \bar{x}_I^{\ddagger},$$

and the existence of such $\bar{x}_I^{\ddagger}$ is guaranteed by $\underline{\theta}_O > F_I^0(0)$.

The above inequality (11) guarantees that there exists a finite moment of time $t$ to satisfy condition (9). Note that $1 - (1-r)r^{r/(1-r)}$ is an increasing function of $r$, converging to 0 as $r \to 0$ and to 1 as $r \to 1$. Therefore, condition (11) is satisfied if $r$ is sufficiently small, namely, if the switching rate is so elastic to payoff difference and thus the difference in switching rates between different types is large enough.

One might think that the escaping dynamic from the "stable" aggregate equilibrium $\bar{x}_I^*$ would be an overshooting due to payoff perturbation or fluctuation in the initial aggregate strategy; but it is *not*. The word "over" would infer so *strong* driving force *toward* an equilibrium that cannot be ceased even when the state reaches the equilibrium. However, as shown analytically in the proof of the above theorem and illustrated numerically in the next example, the aggregate strategy starts *exactly* from $\bar{x}_I^{\ddagger}$, moves *away* from it since the very initial period, and then *monotonically* converges to another aggregate equilibrium $\bar{x}_I = 0$. In short, the cause of non-aggregability in the long-run outcome lies in the distortion of the *direction* of the aggregate transition, not the *strength* of the transition.

Distributional disturbance may push the aggregate strategy out from an aggregate equilibrium that was stable under an aggregable dynamic and it may not return forever, as stated in Corollary 8. Theorem 8 does not tell where the aggregate strategy eventually goes after escaping from an aggregate equilibrium. But, as a binary game is indeed a potential game, Theorem 6 assures that it must converge to either one equilibrium. Thus, if there is only one stable aggregate equilibrium whose $\bar{x}$ is smaller than that of the initial aggregate strategy, the escaping dynamic must converge to this equilibrium. This is summarized in Corollary 3.

**Corollary 3** (Transition from an aggregate equilibrium to another). *Consider the same situation as in Corollary 8. Suppose that the distributional critical mass* $\bar{x}^{\dagger}$ *to decrease* $\bar{x}$ *lies between two isolated aggregate equilibria* $\bar{x}^*$ *and* $\bar{x}^{\flat} < \bar{x}^*$, $\bar{x}^{\flat}$ *is locally stable in the homogenized smooth BRD, and there is no other stable aggregate equilibrium in which the aggregate mass of action-I players is smaller than* $\bar{x}^*$. *Then, if* $\check{Q}_{I,0}^O$ *and* $\check{Q}_{I,0}^I$ *satisfy condition* (9), *then aggregate strategy* $\bar{x}$ *eventually converges to* $\bar{x}^{\flat}$ *after leaving* $\bar{x}^*$.

*Example* 4 (Long-run outcome after escape from $\bar{x} = 0.25$ in Example 1). In Example 1, the sorting pressure in the underlying Bayesian strategy drives out the aggregate strategy from aggregate equilibrium $\bar{x} = 0.25$; thus we could say that this aggregate equilibrium is not stable under the aggregate dynamic in the heterogeneous tBRD, though it is stable under the homogenized smooth BRD and thus under the heterogeneous standard BRD. Figure 1d suggests that the aggregate strategy moves toward $\bar{x} = 0$, though the simulation result on this figure can show the aggregate transition only for a finite length of period. But Corollary 3 assures convergence to $\bar{x} = 0$ in the long run.

Furthermore, if $\bar{x}_0$ in the initial aggregate strategy does not exceed $\bar{x}^\dagger = 0.10$, then the aggregate dynamic of $\bar{x}_t$ never goes beyond this critical mass under the tBRD. As the equilibrium set is globally attracting, this aggregate dynamic must converge to $\bar{x} = 0$. Thus, we could say that $\bar{x} = 0$ is stable even under the aggregate dynamic in the heterogeneous tBRD with the basin of attraction $\bar{x} \in [0, 0.10]$, regardless of the underlying strategy composition. ∎

## 5 Equilibrium selection by distributional stability

In sum, nonaggregability breaks stationarity of aggregate equilibrium and distorts its stability. On the other hand, Bayesian equilibrium possesses stationarity and stability. Especially, in a potential game, the Bayesian strategy is guaranteed to converge to either one Bayesian equilibrium even in nonaggregable dynamics. Correspondingly, the aggregate strategy converges to either one aggregate equilibrium despite of nonaggregable disturbance, though it can be different from the limit state in an aggregable dynamic.

Theorem 2 extends local stability of each (isolated) equilibrium to nonaggregable dynamics. However, it depends on the underlying strategy composition, not only the aggregate strategy, which *local* stable equilibrium to converge. According to Theorem 8, if the strategy composition is largely unsorted, the aggregate strategy may escape even from an aggregate equilibrium that is stable under the homogenized BRD. These positive and negative results on local stability suggest that we can *select* aggregate equilibria by imposing robustness of stability to nonaggregable disturbance to aggregate dynamic at unsorted compositions.

### 5.1 Selection by making the dynamic more sensitive to payoff gains

We expect the transition of aggregate strategy to become more dependent on the strategy composition, as the switching rates become more sensitive to payoff gains from switches. To make a tractable parameterization of payoff sensitivity, consider a tempered BRD with a bounded support of $Q'$: $Q(\breve{\pi})$ increases to $Q(\breve{\pi}^\sharp)$ until $\breve{\pi}$ reaches $\breve{\pi}^\sharp < \infty$ and then keeps $Q(\breve{\pi}^\sharp)$ thereafter, i.e., $Q'(\breve{\pi}) > 0$ for any $\breve{\pi} \in (0, \breve{\pi}^\sharp)$ and $Q(\breve{\pi}) = Q(\breve{\pi}^\sharp)$ for any $\breve{\pi} \geq \breve{\pi}^\sharp$. That is, as raising $\breve{\pi}$, we can gradually get more payoff types to be sensitive in their switching rates to the quantitative degree of payoff gains from switches. Since the payoff sensitivity is the cause of nonaggregability, this parametric increase of payoff sensitivity should strengthen distributional disturbance from the benchmark aggregable dynamic (homogenized BRD) and thus should work to narrow down the set of distributionally stable equilibria. With this parameterization, we can call an aggregate equilibrium the **most robust** to distributional disturbance, if it remains to be the only distributionally stable equilibrium when $\breve{\pi}^\sharp$ increases.

This specification of $Q$ allows to find distributional critical masses just by comparing the payoff difference for the cut-off type $\theta = P_\Theta^{-1}(\bar{x})$ with this upper bound $\breve{\pi}^\sharp$. $\bar{x}^\dagger$ is a distributional critical mass to decrease $\bar{x}$, if

$$P_\Theta^{-1}(\bar{x}^\dagger) - F(\bar{x}^\dagger) \geq \breve{\pi}^\sharp \quad \text{i.e., } P_\Theta^{-1}(\bar{x}^\dagger) \geq F(\bar{x}^\dagger) + \breve{\pi}^\sharp.$$

This condition means that the payoff deficit of the cut-off type is large enough to switch to O at the full rate $Q(\breve{\pi}^\sharp)$. Similarly, $\bar{x}^\ddagger$ is a distributional critical mass to decrease $\bar{x}$, if

$$F(\bar{x}^\ddagger) - P_\Theta^{-1}(\bar{x}^\ddagger) \geq \breve{\pi}^\sharp \quad \text{i.e., } P_\Theta^{-1}(\bar{x}^\ddagger) \leq F(\bar{x}^\dagger) - \breve{\pi}^\sharp.$$

This comparison does not depend on the specification of the upper and lower bounds of the type space. Furthermore, $\breve{\pi}^\sharp$ parameterizes the selection power of this criterion. While $\breve{\pi}^\sharp = 0$ keeps any stable aggregate equilibrium under the homogenized smooth BRD to be nonaggregably stable, less of them will remain as the upper bound $\breve{\pi}^\sharp$ is set higher. Therefore, by raising $\breve{\pi}^\sharp$ high enough, we can generically select a unique aggregate equilibrium.[13]

*Example* 5 (Equilibrium selection in Example 1 revisited again). While we adopted a different tempering function $Q(\breve{\pi}) = \breve{\pi}^3$ in Example 1 so to easily find an escaping path from $\bar{x} = 0.25$, we could adopt $Q$ as suggested above. From Figure 6, we can find the greatest payoff deficit of action O is $1/1600$ at $\bar{x} = 0.225$, which is in the basin of attraction to $\bar{x} = 0.25$ under the homogenized BRD. Therefore, both $\bar{x} = 0$ and $\bar{x} = 0.25$ are nonaggregably stable if $\breve{\pi}^\sharp < 1/1600$. But, if $\breve{\pi}^\sharp$ exceeds $1/1600$, then aggregate equilibrium $\bar{x} = 0.25$ has no distributional critical mass in $(0.2, 0.25)$ to increase $\bar{x}$ toward this aggregate equilibrium and thus $\bar{x} = 0.25$ is no longer nonaggregably stable. On the other hand, $\bar{x} = 0$ is still nonaggregably stable, because it still has distributional critical masses in $(0, 0.2)$ to decrease $\bar{x}$ toward $\bar{x} = 0$ remains to exist. Thus, our selection criterion chooses aggregate equilibrium $\bar{x} = 0$ as the most robust equilibrium. ∎

## 5.2 Relation with risk dominance

Consider a linear aggregate game such as $F(\bar{x}) := (1-c)\bar{x} - c(1-\bar{x})$ with $c > 0$. This can be seen as random matching in a normal-form coordination game. Here $-c < 0$ is the payoff from $I$ when nobody takes $I$ and $1 - c > 0$ is the payoff from $I$ when everybody takes $I$. Without heterogeneity $\theta \equiv 0$, there are three Nash equilibria: $\bar{x} = 0, 1, a/(a+b)$. $x = 1$ is risk dominant, if $I$ is optimal when $\bar{x} = 1/2$, i.e., if $c < 1/2$; $x = 0$ is risk dominant, if $O$ is optimal when $\bar{x} = 1/2$, i.e., if $c > 1/2$.

---

[13]There is an exception. Notice that distributional critical masses for a nonaggregably stable aggregate equilibrium are located in the basin of attraction to this aggregate equilibrium under the homogenized smooth BRD. Then we can search for distributional critical masses from this basin of attraction by checking the payoff deficit of the cut-off type. Thus, when selecting equilibrium, we compare the greatest payoff deficit in the basin of attraction to each aggregate equilibrium under the homogenized smooth BRD. More precisely, for each aggregate equilibrium, we should take the smaller of the greatest deficit in the part of the basin of attraction where $\bar{x}$ is smaller than (i.e., located left to) the equilibrium (for a distributional critical mass to increase $\bar{x}$ toward this equilibrium) and the greatest one in the right part of the basin of attraction (for a distributional critical mass to decrease $\bar{x}$). If the greatest payoff deficits in the basins of attraction to several aggregate equilibria happen to be equal, our selection criterion cannot choose a unique equilibrium with a bounded tempering function $Q$.

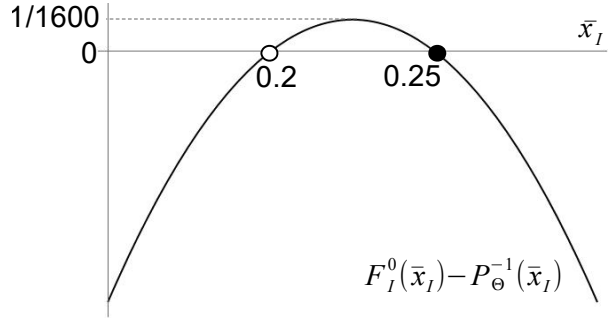

*Figure 6:* Payoff difference for the cut-off type $\theta = P_\Theta^{-1}(\bar{x})$ in Example 1.

To introduce heterogeneity, assume that $P_\Theta(0) = 1/2$ and $P_\Theta$ is point-symmetric around $\theta = 0$. Each equilibrium is shifted when heterogeneity is introduced. Then, under tBRD with a bounded support of $Q'$, the most robust aggregate equilibrium is the one shifted from the risk dominant equilibrium in the normal-form coordination game.

# 6 Concluding remarks

In a binary-choice game—or more generally in a potential game, a Bayesian equilibrium is asymptotically stable commonly under any admissible dynamics, if and only if an aggregate equilibrium is asymptotically stable in the homogenized BRD where persistent payoff heterogeneity averages to homogeneous transitory payoff shocks. However, nonaggregable dynamics may leave the aggregate equilibrium and converge to another equilibrium, when the initial Bayesian strategy is far from the Bayesian equilibrium and thus it is out of the basin of attraction under the heterogeneous dynamic. We define distributional stability of an aggregate equilibrium by imposing robustness of local stability to distributional disturbance so as to guarantee convergence to the aggregate equilibrium from any nearby aggregate strategy regardless of the underlying Bayesian strategy in a given nonaggregable dynamic. Under aggregagble dynamics, it is equivalent to local stability under the homogenized dynamic; however distributional stability is stronger than local stability under nonaggregable dynamics. We can narrow down a set of stable equilibria from the one under aggregable dynamics to the one under nonaggregate dynamics. This is the basic idea for our equilibrium selection under deterministic heterogeneous dynamics.

In our equilibrium selection by deterministic heterogeneous dynamics, we check the distributional robustness of local stability by starting the dynamic from a completely reversed strategy composition in which the initial choices of agents are selected *against* their incentives while keeping the aggregate strategy. It may be interpreted as testing the aggregate dynamic by the "worst case" about estimation of the underlying strategy composition behind the given aggregate strategy. From this worst case, we let agents adjust their choices rationally (but myopically) and see whether they will return to the aggregate equilibrium or not.

The idea of testing robustness by starting from an extreme case on possible payoffs and check-ing cascade effects might share the same spirit as equilibrium selection by global game, though the latter essentially assumes equilibrium behavior. It would be a next agenda to rigorously relate distributional stability with global game and to see if heterogeneous dynamics may or may not give evolutionary explanation to equilibrium selection by global games.

What's a rational to consider a reversed composition? We could imagine rare stochastic shuffle of agents' types, happening independently of their choices of actions; it only shuffles individual agents' payoff types while keeping the distribution of payoff types among them. The shuffle of payoff types alone does not alter the aggregate strategy but changes the strategy composition. Even from a Bayesian equilibrium, there is a positive probability to change it to a reversed composition. If our deterministic dynamic works as the mean dynamic in such a stochastic environment, we can expect the aggregate strategy to be driven out from a distributionally instable equilibrium once such a catastrophic shuffle occurs. On the other hand, if the initial aggregate strategy is in the distributional basin of attraction to a distributionally stable equilibrium, the shuffle cannot take the aggregate strategy away from the equilibrium. We could conjecture that distributional stability may be equivalent to stochastic stability under such shuffles of payoff types, while it adds to the agenda of the next research as well.[14]

# References

BILLINGSLEY, P. (1979): *Probability and Measure*. John Wiley & Sons, New York, 1 edn.

ELY, J., AND W. H. SANDHOLM (2005): "Evolution in Bayesian Games I: Theory," *Games and Economic Behavior*, 53, 83–109.

GILBOA, I., AND A. MATSUI (1991): "Social Stability and Equilibrium," *Econometrica*, 59(3), 859–867.

HOFBAUER, J. (1995): "Stability for the best response dynamics," mimeo, University of Vienna.

KUZMICS, C. (2011): "On the Elimination of Dominated Strategies in Stochastic Models of Evolution with Large Populations," *Games and Economic Behavior*, 72(2), 452–66.

MILGROM, P. R., AND R. J. WEBER (1985): "Distributional strategies for games with incomplete information," *Mathematics of operations research*, 10(4), 619–632.

NORMAN, T. W. (2009): "Rapid evolution under inertia," *Games and Economic Behavior*, 66(2), 865–879.

——— (2010): "Cycles versus equilibrium in evolutionary games," *Theory and Decision*, 69(2), 167–182.

---

[14]Kuzmics (2011) and Norman (2009, 2010) introduce switching costs to stochastic evolution with noise.

SANDHOLM, W. H. (2007): "Evolution in Bayesian games II: Stability of purified equilibria," *Journal of Economic Theory*, 136(1), 641–667.

ZUSAI, D. (2018a): "Nonaggregate evolutionary dynamics under heterogeneity in payoffs and strategy revision protocols," mimeo, Temple University.

——— (2018b): "Tempered best response dynamics," *International Journal of Game Theory*, 47(1), 1–34.
26

# A Proofs

## A.1 Proof of Theorem 3

We prove part i) of the theorem. First of all, let $x$ be an arbitrary Bayesian strategy with $\mathbb{E}_\Theta x = \bar{x}^\dagger$. Then, the transition of the aggregate strategy is

$$\mathbb{E}_\Theta v^F[x] = \int_{-\infty}^{F(\bar{x}^\dagger)} Q_{OI}(F(\bar{x}^\dagger), \theta)\,(1 - x(\theta))dP_\Theta(\theta) - \int_{F(\bar{x}^\dagger)}^{+\infty} Q_{IO}(F(\bar{x}^\dagger), \theta)\,x(\theta)dP_\Theta(\theta).$$

Specifically, let $x^\dagger$ be the *perfectly sorted* Bayesian strategy such that $x^\dagger(\theta) = \mathbf{1}(\theta \leq P_\Theta^{-1}(\bar{x}^\dagger))$. Then, the transition of the aggregate strategy is

$$\mathbb{E}_\Theta v^F[x^\dagger] = -\int_{F(\bar{x}^\dagger)}^{P_\Theta^{-1}(\bar{x}^\dagger)} Q_{IO}(F(\bar{x}^\dagger), \theta) \cdot 1 dP_\Theta(\theta) < 0,$$

since $F(\bar{x}^\dagger) < P_\Theta^{-1}(\bar{x}^\dagger)$ and $Q_{IO}(\pi) > 0$ whenever $\pi_O > \pi_I$.

Comparison of these two equations implies

$$\mathbb{E}_\Theta v^F[x] - \mathbb{E}_\Theta v^F[x^\dagger] = \int_{-\infty}^{F(\bar{x}^\dagger)} Q_{OI}(F(\bar{x}^\dagger), \theta)\,(1 - x(\theta))dP_\Theta(\theta)$$

$$+ \int_{F(\bar{x}^\dagger)}^{P_\Theta^{-1}(\bar{x}^\dagger)} Q_{IO}(F(\bar{x}^\dagger), \theta)\,(1 - x(\theta))dP_\Theta(\theta)$$

$$- \int_{P_\Theta^{-1}(\bar{x}^\dagger)}^{+\infty} Q_{IO}(F(\bar{x}^\dagger), \theta)\,x(\theta)dP_\Theta(\theta)$$

$$\leq Q_{IO}(F(\bar{x}^\dagger), P_\Theta^{-1}(\bar{x}^\dagger)) \left( \bar{x}^\dagger - \int_{-\infty}^{P_\Theta^{-1}(\bar{x}^\dagger)} x(\theta)dP_\Theta(\theta) \right)$$

$$- Q_{IO}(F(\bar{x}^\dagger), P_\Theta^{-1}(\bar{x}^\dagger)) \int_{P_\Theta^{-1}(\bar{x}^\dagger)}^{+\infty} x(\theta)dP_\Theta(\theta)$$

$$= Q_{IO}(F(\bar{x}^\dagger), P_\Theta^{-1}(\bar{x}^\dagger))\,(\bar{x}^\dagger - \bar{x}^\dagger) = 0.$$

The inequality comes from (3). Therefore, we have

$$\mathbb{E}_\Theta v^F[x] \leq \mathbb{E}_\Theta v^F[x^\dagger] < 0.$$

Part ii) is proven by a similar comparison between $\mathbb{E}_\Theta v^F[x]$ of an arbitrary Bayesian strategy and $\mathbb{E}_\Theta v^F[x^\ddagger]$ of the sorted Bayesian strategy $x^\ddagger$ defined in the same way as $x^\dagger$.

## A.2 Proof of Theorem 4

Under the tempered BRD in the binary game, the composite dynamic reduces to

$$\dot{x}(\theta) = v^F[\bar{x}](\theta) = \begin{cases} Q(F(\bar{x}) - \theta)(1 - x(\theta)) & \text{if } F(\bar{x}) > \theta, \\ 0 & \text{if } F(\bar{x}) = \theta, \\ -Q(\theta - F(\bar{x}))x(\theta) & \text{if } F(\bar{x}) < \theta, \end{cases}$$

and thus the aggregate dynamic reduces to

$$\dot{x} = \bar{V}(x) = \int_{-\infty}^{F(\bar{x})} Q(F(\bar{x}) - \theta)(1 - x(\theta))d\mathbb{P}_\Theta(\theta) + \int_{F(\bar{x})}^{\infty} \{-Q(\theta - F(\bar{x}))x(\theta)\}d\mathbb{P}_\Theta(\theta),$$

where $\bar{x} = \mathbb{E}_\Theta x$.

First, we decompose the aggregate dynamic $\bar{V}$ into $\bar{V} = \bar{V}_0 + \Delta\bar{V}_1 + \Delta\bar{V}_2 + \Delta\bar{V}_3$ with

$$\bar{V}_0(x) := \int_{-\infty}^{\theta^*} Q(\theta^* - \theta)(1 - x(\theta))d\mathbb{P}_\Theta(\theta) + \int_{\theta^*}^{+\infty} \{-Q(\theta - \theta^*)x(\theta)\}d\mathbb{P}_\Theta(\theta),$$

$$\Delta\bar{V}_1(x) = \int_{-\infty}^{\theta^*} \{Q(F(\bar{x}) - \theta) - Q(\theta^* - \theta)\}(1 - x(\theta))d\mathbb{P}_\Theta(\theta),$$

$$\Delta\bar{V}_2(x) = \int_{\theta^*}^{F(\bar{x})} \{Q(F(\bar{x}) - \theta)(1 - x(\theta)) + Q(\theta - \theta^*)x(\theta)\}d\mathbb{P}_\Theta(\theta),$$

$$\Delta\bar{V}_3(x) = -\int_{F(\bar{x})}^{+\infty} \{Q(\theta - F(\bar{x})) - Q(\theta - \theta^*)\}x(\theta)d\mathbb{P}_\Theta(\theta).$$

With $\bar{x}^* =: \mathbb{E}_\Theta x^*$ and $\theta^* := F(\bar{x}^*)$, we have the linear approximation of $Q$ as

$$Q(F(\bar{x}) - \theta) = Q(\theta^* - \theta) + Q'(\theta^* - \theta)F'(\bar{x}^*)(\bar{x} - \bar{x}^*) + o(\bar{x} - \bar{x}^*),$$

where $o$ is Landau's little-o, i.e., $|o(\delta)/\delta| \to 0$ as $\delta \to 0$. It follows that

$$\Delta\bar{V}_1(x) = F'(\bar{x}^*)(\bar{x} - \bar{x}^*) \left\{ \int_{-\infty}^{\theta^*} Q'(\theta^* - \theta)(1 - x(\theta))d\mathbb{P}_\Theta(\theta) + o(\bar{x} - \bar{x}^*) \right\}.$$

Similarly, we obtain

$$\Delta\bar{V}_3(x) = F'(\bar{x}^*)(\bar{x} - \bar{x}^*) \left\{ \int_{F(\bar{x})}^{+\infty} Q'(\theta - \theta^*)x(\theta)d\mathbb{P}_\Theta(\theta) + o(\bar{x} - \bar{x}^*) \right\}.$$

Presume that $F(\bar{x}) \geq \theta^*$. Then, $\Delta\bar{V}_2(x)$ is $o(\bar{x}^* - \bar{x})$, since $\Delta V_2(x) \geq 0$ and

$$\Delta\bar{V}_2(x) \leq \int_{\theta^*}^{F(\bar{x})} \{Q(F(\bar{x}) - \theta^*)(1 - x(\theta)) + Q(F(\bar{x}) - \theta^*)x(\theta)\}dP_\Theta(\theta)$$

$$= Q(F(\bar{x}) - \theta^*)\}\mathbb{P}_\Theta((\theta^*, F(\bar{x})))$$

$$\leq Q(F(\bar{x}) - \theta^*)\bar{p}_\Theta(F(\bar{x}) - \theta^*)$$

$$= \{Q'(0) + F'(\bar{x}^*)(\bar{x}^* - \bar{x}) + o(\bar{x}^* - \bar{x})\}\bar{p}_\Theta\{(F'(\bar{x}^*)(\bar{x}^* - \bar{x}) + o(\bar{x}^* - \bar{x})\}$$

$$= Q'(0)F'(\bar{x}^*)^2\bar{p}_\Theta(\bar{x}^* - \bar{x})^2 + o((\bar{x}^* - \bar{x})^2).$$

In the first inequality, we use the assumption that $Q$ is increasing and $F(\bar{x}) \geq \theta^*$. For the second equality, notice that, in both two parts, $\bar{x}^*$ is an aggregate equilibrium and thus $\theta^* = F(\bar{x}^*)$.

*Proof of part 1.* First, since $x^*$ is not a Bayesian equilibrium but $\bar{x}^*$ is an aggregate equilibrium, the participant composition $X^*$ should satisfy

$$\mathbb{P}_\Theta(\theta^*) - X^*((-\infty, \theta^*]) = X^*((\theta^*, +\infty)) > 0.$$

Choose $\varepsilon \in (0, \bar{\varepsilon})$ arbitrarily. Define Bayesian strategy $x^\dagger$ as

$$x^\dagger(\theta) := \begin{cases} (1-\varepsilon)x^*(\theta) + \varepsilon & \text{if } \theta < \theta^*, \\ x^*(\theta) & \text{otherwise.} \end{cases}$$

This Bayesian strategy induces the aggregate participation rate $\bar{x}^\dagger$ such that

$$\bar{x}^\dagger := \mathbb{E}_\Theta x^\dagger = \mathbb{E}_\Theta x^* + \int_{-\infty}^{\theta^*} \varepsilon(1 - x^*(\theta))d\mathbb{P}_\Theta(\theta)$$

$$= \bar{x}^* + \varepsilon \underbrace{\{\mathbb{P}_\Theta(\theta^*) - X^*((-\infty, \theta^*])\}}_{A > 0} > \bar{x}^*.$$

Observe that

$$\bar{V}_0(x^\dagger) = \int_{-\infty}^{\theta^*} Q(\theta^* - \theta)(1 - x^\dagger(\theta))d\mathbb{P}_\Theta(\theta) \int_{\theta^*}^{+\infty} \{-Q(\theta - \theta^*)x^\dagger(\theta)\}d\mathbb{P}_\Theta(\theta)$$

$$= V(x^*) + \int_{-\infty}^{\theta^*} Q(\theta^* - \theta)(x^*(\theta) - x^\dagger(\theta))d\mathbb{P}_\Theta(\theta) + \int_{\theta^*}^{+\infty} Q(\theta - \theta^*)\{x^*(\theta) - x^\dagger(\theta)\}d\mathbb{P}_\Theta(\theta)$$

$$= \int_{-\infty}^{\theta^*} Q(\theta^* - \theta)\varepsilon(1 - x^*(\theta))d\mathbb{P}_\Theta(\theta)$$

$$= \varepsilon \underbrace{\int_{-\infty}^{\theta^*} Q(\theta^* - \theta)(1 - x^*(\theta))d\mathbb{P}_\Theta(\theta)}_{B > 0} > 0.$$

In the second equality, we use the assumption that $x^*$ satisfies the balancing condition and thus $\bar{V}(x^*) = 0$. The last inequality comes from the fact that $\mathbb{P}_\Theta(\theta^*) - X^*((-\infty, \theta^*]) > 0$ and thus $1 - x^*(\theta) > 0$ in a positive measure subset of $(-\infty, \theta^*]$.

We have

$$\int_{-\infty}^{\theta^*} Q'(\theta^* - \theta)(1 - x^\dagger(\theta))d\mathbb{P}_\Theta(\theta) = (1 - \varepsilon)\underbrace{\int_{-\infty}^{\theta^*} Q'(\theta^* - \theta)(1 - x^*(\theta))d\mathbb{P}_\Theta(\theta)}_{C \geq 0} \geq 0,$$

since $Q' \geq 0$ and $1 \geq x^*(\theta)$. Thus,

$$\Delta V_1(x) = F'(\bar{x}^*)A\varepsilon\{(1-\varepsilon)C + o(A\varepsilon)\} = F'(\bar{x}^*)AC\varepsilon + o(\varepsilon).$$

We have

$$\int_{F(\bar{x}^\dagger)}^{+\infty} Q'(\theta - \theta^*)x^\dagger(\theta)d\mathbb{P}_\Theta(\theta) = \int_{F(\bar{x}^\dagger)}^{+\infty} Q'(\theta - \theta^*)x^*(\theta)d\mathbb{P}_\Theta(\theta)$$

$$= \underbrace{\int_{\theta^*}^{+\infty} Q'(\theta - \theta^*)x^*(\theta)d\mathbb{P}_\Theta(\theta)}_{D \geq 0} - \int_{\theta^*}^{F(\bar{x}^\dagger)} Q'(\theta - \theta^*)x^*(\theta)d\mathbb{P}_\Theta(\theta).$$

The latter integral is non-negative and bounded by a linear function of $\varepsilon$, because

$$\int_{\theta^*}^{F(\bar{x}^\dagger)} Q'(\theta - \theta^*)x^*(\theta)d\mathbb{P}_\Theta(\theta) \leq \bar{Q}'\bar{p}_\Theta(F(\bar{x}^\dagger) - \theta^*) = \bar{Q}'\bar{p}_\Theta(A\varepsilon + o(\varepsilon)).$$

Here $\bar{Q}'$ is the maximum of $Q'(q)$ in $[0, F(\bar{x}^\dagger) - \theta^*]$; recall that $Q$ is continuously differentiable and

$F(\bar{x}^\dagger) > F(\bar{x}^*) = \theta^*$ by $dF/d\bar{x} > 0$. Hence,

$$\Delta V_3(x) = F'(\bar{x}^*)A\varepsilon \left\{ D - \int_{\theta^*}^{F(\bar{x}^\dagger)} Q'(\theta - \theta^*)x^*(\theta)d\mathbb{P}_\Theta(\theta) + o(A\varepsilon) \right\}$$

$$= F'(\bar{x}^*)AD\varepsilon + o(\varepsilon).$$

Therefore, we have

$$\bar{V}(x^\dagger) = \{B + F'(\bar{x}^*)A(C + D)\}\varepsilon + o(\varepsilon)$$

When $\varepsilon$ is sufficiently small, $\dot{\bar{x}} = \bar{V}(x^\dagger)$ is positive. $\square$

*Proof of part 2.* Let $e \in \mathbb{R}_+$ be small enough to meet $e < \min\{1, \bar{q}/4\}$ and $F(\bar{x}^*) - 4e, F(\bar{x}^*) + e \in \Theta$. Choose $w \in \mathbb{R}_+$ arbitrarily such that

$$\int_e^{2e} Q(q)dq \bigg/ \int_{3e}^{4e} Q(q)dq < w < 1.$$

As $Q$ is increasing in $[0, \bar{q}]$ and $4e < \bar{q}$, the fraction on the LHS is smaller than 1. Then, choose $\varepsilon \in (0, \bar{\varepsilon})$ so small that $X^\dagger$ defined from density $x^\dagger$ given below satisfies $F(\mathbb{E}_\Theta x^\dagger) < \theta^* + e$, $x^\dagger(\theta) \in (0, 1)$ for all $\theta$, and $d(X^\dagger, X^*) < \varepsilon$:

$$x^\dagger(\theta) := \begin{cases} x^*(\theta) + \varepsilon/p_\Theta(\theta) = \varepsilon/p_\Theta(\theta) & \text{if } \theta \in [\theta^* + e, \theta^* + 2e), \\ x^*(\theta) - w\varepsilon/p_\Theta(\theta) = 1 - w\varepsilon/p_\Theta(\theta) & \text{if } \theta \in [\theta^* - 4e, \theta^* - 3e), \\ x^*(\theta) & \text{otherwise.} \end{cases}$$

Note that, for arbitrary $B \in \mathcal{B}_\mathbb{R}$,

$$X^\dagger(B) = \int_B x^\dagger(\theta)p_\Theta(\theta)d\theta$$

$$= X^*(B) + \varepsilon \int_{B \cap [\theta^*+e, \theta^*+2e)} \frac{\varepsilon}{p_\Theta(\theta)}p_\Theta(\theta)d\theta + \varepsilon \int_{B \cap [\theta^*-4e, \theta^*-3e)} \frac{w\varepsilon}{p_\Theta(\theta)}p_\Theta(\theta)d\theta$$

$$\in [X^*(B) - we\varepsilon, X^*(B) + \varepsilon] \subset [X^*(B) - \varepsilon, X^*(B) + \varepsilon].$$

The last line comes from $e \leq 1$ and $w < 1$. Therefore, we have $X^*(B) \leq X^\dagger(B) + \varepsilon \leq X^\dagger(B^\varepsilon) + \varepsilon$ and $X^\dagger(B) \leq X^*(B) + \varepsilon \leq X^*(B^\varepsilon) + \varepsilon$ for any $B \in \mathcal{B}_\mathbb{R}$: namely, $d_\mathcal{M}(X^\dagger, X^*) < \varepsilon$.

This Bayesian strategy induces the aggregate participation rate $\bar{x}^\dagger$ such that

$$\bar{x}^\dagger := \mathbb{E}_\Theta x^\dagger = \mathbb{E}_\Theta x^* + \int_{\theta^*+e}^{\theta^*+2e} \frac{\varepsilon}{p_\Theta(\theta)}d\mathbb{P}_\Theta(\theta) - \int_{\theta^*-4e}^{\theta^*-3e} \frac{w\varepsilon}{p_\Theta(\theta)}d\mathbb{P}_\Theta(\theta)$$

$$= \bar{x}^* + \underbrace{(1-w)e}_{A'>0}\varepsilon > \bar{x}^*.$$

Furthermore, we have

$$\bar{V}_0(x^\dagger) = \int_{\underline{\theta}}^{\theta^*} Q(\theta^* - \theta)(1 - x^\dagger(\theta))d\mathbb{P}_\Theta(\theta) + \int_{\theta^*}^{\bar{\theta}} \{-Q(\theta - \theta^*)x^\dagger(\theta)\}d\mathbb{P}_\Theta(\theta)$$

$$= \int_{\theta^*-4e}^{\theta^*-3e} Q(\theta^* - \theta)\frac{w\varepsilon}{p_\Theta(\theta)}d\mathbb{P}_\Theta(\theta) - \int_{\theta^*+e}^{\theta^*+2e} Q(\theta - \theta^*)\frac{\varepsilon}{p_\Theta(\theta)}d\mathbb{P}_\Theta(\theta)$$

$$= \varepsilon' \underbrace{\left( w \int_{3e}^{4e} Q(q)dq - \int_{e}^{2e} Q(q)dq \right)}_{B' > 0} > 0$$

The definition of $x^\dagger$ implies

$$\int_{-\infty}^{\theta^*} Q'(\theta^* - \theta)(1 - x^\dagger(\theta))d\mathbb{P}_\Theta(\theta)$$

$$= \int_{\theta^* - 4e}^{\theta^* - 3e} Q'(\theta^* - \theta) \frac{w\varepsilon}{p_\Theta(\theta)} d\mathbb{P}_\Theta(\theta) = w\varepsilon \int_{\theta^* - 4e}^{\theta^* - 3e} Q'(\theta^* - \theta) d\theta$$

$$= \varepsilon \underbrace{w(Q(4e) - Q(3e))}_{C' > 0} > 0.$$

It follows that

$$\Delta V_1(x) = F'(\bar{x}^*)A'\varepsilon\{\varepsilon C' + o(A'\varepsilon)\}$$
$$= F'(\bar{x}^*)A'C'\varepsilon^2 + o(\varepsilon^2) = o(\varepsilon).$$

Similarly, we have

$$\int_{F(\bar{x}^\dagger)}^{+\infty} Q'(\theta - \theta^*)x^\dagger(\theta)d\mathbb{P}_\Theta(\theta)$$

$$= \int_{\theta^* + e}^{\theta^* + 2e} Q'(\theta - \theta^*) \frac{\varepsilon}{p_\Theta(\theta)} d\mathbb{P}_\Theta(\theta) = \varepsilon \int_{\theta^* + e}^{\theta^* + 2e} Q'(\theta - \theta^*) d\theta$$

$$= \varepsilon \underbrace{(Q(2e) - Q(e))}_{D' > 0} > 0$$

Note that $\theta^* + e > F(\bar{x}^\dagger)$. Hence,

$$\Delta V_3(x^\dagger) = F'(\bar{x}^*)A'\varepsilon(\varepsilon D' + o(A'\varepsilon))$$
$$= F'(\bar{x}^*)A'D'\varepsilon^2 + o(\varepsilon^2) = o(\varepsilon).$$

Therefore, we have

$$\bar{V}(x^\dagger) = B'\varepsilon + o(\varepsilon).$$

Hence, when $\varepsilon^\dagger$ is sufficiently small, $\dot{x} = \bar{V}(x^\dagger)$ is positive. □

## A.3 Proof of Theorem 5

*Proof.* For each $a \in \mathcal{A}$, the time derivative $\dot{x}_a$ is

$$\dot{x}_{a,0} = \mathbb{E}_\Theta \dot{x}_{a,0} = \int_\Theta \dot{x}_{a,0} \mathbb{P}_\Theta(d\theta)$$

$$= \int_{\beta_a^{-1}(\bar{x}^*)} \sum_{i \in \mathcal{A}\setminus\{a\}} Q_{ia}(F(\bar{x}^*), \theta(\omega))x_0(\theta)\mathbb{P}_\Theta(d\theta)$$

$$- \int_{\Theta\setminus b_a^{-1}(\bar{x}^*)} \sum_{j \in \mathcal{A}\setminus\{a\}} Q_{aj}(F(\bar{x}^*), \theta(\omega))x_{a,0}(d\theta)\mathbb{P}_\Theta(d\theta)$$

$$= \int_{-\infty}^{\infty} q_{a,0}^I \check{Q}_{a,0}^I(dq_{a,0}^I) - \int_{-\infty}^{\infty} q_{a,0}^O \check{Q}_{a,0}^O(dq_{a,0}^O)$$

$\square$

## A.4 Proof of Theorem 6

Fix time $T \in [0, \Delta]$ arbitrarily and presume that the aggregate state is $\bar{x}^*$ at time $T$. For each pair of two actions $(i, j) \in \mathcal{A}^2$, let $\Omega_{ij,T}$ be the mass of agents who have been playing action $i$ at time $T$ but has action $j$ as the *unique* best response to the aggregate state $\bar{x}^*$:

$$\Omega_{ij,T} := \{\omega \in \Omega : \mathfrak{a}_T(\omega) = i \text{ and } \theta(\omega) \in \beta_j^{-1}(\bar{x}^*)\}.$$

Assumption 2 implies that $\mathbb{P}_\Omega$-almost every type has a unique best response and thus $\mathbb{P}_\Omega \left( \bigcup_{(i,j) \in \mathcal{A}^2} \Omega_{ij,T} \right) = \mathbb{P}_\Omega(\Omega) = 1$. Define function $q_{ij,T} : \Omega_{ij,T} \to \mathbb{R}$ by

$$q_{ij,T}(\omega) = Q_{ij}(F(\bar{x}^*), \theta(\omega)) \qquad \text{for each } \omega \in \Omega_{ij,T}.$$

This is the switching rate for agent $\omega$ at time $T$ from action $i$ to action $j$.

For each $a \in \mathcal{A}$, define $\Omega_{a,T}^I, \Omega_{a,T}^O \subset \Omega$ by

$$\Omega_{a,T}^I := \bigcup_{i \in \mathcal{A} \setminus \{a\}} \Omega_{ia,T}, \qquad \Omega_{a,T}^O := \bigcup_{j \in \mathcal{A} \setminus \{a\}} \Omega_{aj,T},$$

and functions $q_{a,T}^I : \Omega_{a,T}^I \to \mathbb{R}$ and $q_{a,T}^O : \Omega_{a,T}^O \to \mathbb{R}$ by

$$q_{a,T}^I(\omega) = q_{ia,T}(\omega) \qquad \text{whenever } \omega \in \Omega_{ia,T} \subset \Omega_a^I,$$
$$q_{a,T}^O(\omega) = q_{aj,T}(\omega) \qquad \text{whenever } \omega \in \Omega_{aj,T} \subset \Omega_a^O.$$

These are agent $\omega$'s switching rate to its best response at time $T$. Notice that $\{\Omega_{ij,T}\}_{(i,j) \in \mathcal{A}^2}$ has no overlap, by definition: $\Omega_{ij,T} \cap \Omega_{i'j',T} = \emptyset$ whenever $(i,j) \neq (i',j')$.

We see $q_{a,T}^I$ and $q_{a,T}^O$ respectively as random variables over $\Omega_{a,T}^I$ and $\Omega_{a,T}^O$. We can construct their cumulative distribution functions from the composition $\mathbf{X}_T$: for each $a \in \mathcal{A}$ and $q \in \bar{\mathbb{R}} := \mathbb{R} \cup \{-\infty, +\infty\}$,

$$\check{Q}_{a,T}^I(q) = \mathbb{P}_\Omega \left( \{\omega \in \Omega_{a,T}^I : q_{a,T}^I(\omega) \leq q\} \right) = \sum_{i \in \mathcal{A} \setminus \{a\}} \mathbb{P}_\Omega \left( \{\omega \in \Omega_{ia,T} : q_{ia,T}(\omega) \leq q\} \right)$$

$$= \sum_{i \in \mathcal{A} \setminus \{a\}} \mathbb{P}_\Omega \left( \{\omega \in \Omega : Q_{ia}(F(\bar{x}^*), \theta(\omega)) \leq q \text{ and } \mathfrak{a}_T(\omega) = i \text{ and } \theta(\omega) \in \beta_a^{-1}(\bar{x}^*)\} \right)$$

$$= \sum_{i \in \mathcal{A} \setminus \{a\}} x_T \left( \{\theta \in \Theta : Q_{ia}(F(\bar{x}^*), \theta) \leq q \text{ and } \theta \in \beta_a^{-1}(\bar{x}^*)\} \right)$$

$$= \sum_{i \in \mathcal{A} \setminus \{a\}} \mathbb{E}_\Theta \left[ \mathbf{1}\left(\theta \in \beta_a^{-1}(\bar{x}^*)\right) \mathbf{1}\left(Q_{ia}(F(\bar{x}^*), \theta) \leq q\right) x_T(\theta) \right],$$

$$\check{Q}_{a,T}^O(q) = \mathbb{P}_\Omega \left( \{\omega \in \Omega_{a,T}^O : q_{a,T}^O(\omega) \leq q\} \right) = \sum_{j \in \mathcal{A} \setminus \{a\}} \mathbb{P}_\Omega \left( \{\omega \in \Omega_{aj,T} : q_{aj,T}(\omega) \leq q\} \right)$$

$$= \sum_{j \in \mathcal{A} \setminus \{a\}} \mathbb{P}_\Omega \left( \{\omega \in \Omega : Q_{aj}(F(\bar{x}^*), \theta(\omega)) \leq q \text{ and } \mathfrak{a}_T(\omega) = a \text{ and } \theta(\omega) \in \beta_j^{-1}(\bar{x}^*)\} \right)$$

$$= \sum_{j \in \mathcal{A} \setminus \{a\}} X_{a,T}\left(\{\theta \in \Theta \,:\, Q_{aj}(F(\bar{x}^*), \theta) \leq q \text{ and } \theta \in \beta_j^{-1}(\bar{x}^*)\}\right)$$

$$= \sum_{j \in \mathcal{A} \setminus \{a\}} \mathbb{E}_\Theta \left[ \mathbf{1}\left(\theta \in \beta_j^{-1}(\bar{x}^*)\right) \mathbf{1}\left(Q_{aj}(F(\bar{x}^*), \theta) \leq q\right) x_{j,T}(\theta) \right].$$

With $T = 0$, these are consistent with the definitions of $\check{Q}^I_{a,0}, \check{Q}^O_{a,0}$ in (5). Note that we could replace $\beta^{-1}$ with $b^{-1}$ without changing these measures, thanks to Assumption 2.

*Proof of the "if" part in Theorem 6.* The stationarity condition (7) implies that the two integrals in (6) are equal and thus $\dot{x}_{a,0} = 0$.

$\square$

In the proof of the "only if" part, we use the following lemma.

**Lemma 1.** *Let a random variable $\kappa^i$ ($i = 1, 2$) have a continuous c.d.f. $G^i$ with density $g^i$. Besides, assume that a function $\beta : \mathbb{R} \to [0, \bar{b}] \subset \mathbb{R}$ be nondecreasing and especially strictly increasing in an interval $\bar{K} \subset \mathbb{R}$ with $\inf\{\beta(k) | k \in \bar{K}\} = 0$ and $\sup\{\beta(k) | k \in \bar{K}\} = \bar{b} > 0$. Suppose that there exists $\epsilon > 0$ such that*

$$\int_{-\infty}^{+\infty} \exp(-\beta(k)\tau) g^1(k) dk = \int_{-\infty}^{+\infty} \exp(-\beta(k)\tau) g^2(k) dk \quad \text{for all } \tau \in (-\epsilon, \epsilon). \tag{A.1}$$

*Then, we have*

$$g^1(k) = g^2(k) \quad \text{for all } k \in \bar{K}.$$

*Proof of Lemma 1.* Define function $B^i : \mathbb{R} \to \mathbb{R}$ as $B^i(b) := \mathbb{P}[\beta(\kappa^i) \leq b] = G^i(\sup \beta^{-1}(b))$ ($i = 1, 2$). Then, it is the cumulative distribution function of the random variable $\beta(\kappa^i)$ with a bounded support, $\operatorname{supp}(B^i) = [0, \bar{b}]$. Each side of the assumption (A.1) is

$$\int_{-\infty}^{\infty} \exp(-\beta(k)\tau) g^i(k) dk = \int_0^{\bar{b}} e^{-b\tau} dB^i(b),$$

namely the moment generating function of each distribution function $B^i$.

So the assumption (A.1) means that these two moment generating functions coincide with each other for $\tau \in (-\epsilon, \epsilon)$. As these two have bounded support $[0, \bar{b}]$, this implies their identity $B^1(b) = B^2(b)$ for all $b \in \mathbb{R}$ (Billingsley, 1979, p.253). $B^1 \equiv B^2$ means

$$G^1(\beta^{-1}(b)) = G^2(\beta^{-1}(b)) \quad \text{for all } b \in [0, \bar{b}];$$

because $\beta$ is non-decreasing and strictly increasing in $\bar{K}$, it is equivalent to

$$G^1(k) = G^2(k), \quad \text{i.e., } g^1(k) = g^2(k) \quad \text{for all } k \in \bar{K}.$$

$\square$

*Proof of the "only if" part in Theorem 6.* As long as the aggregate state stays at $\bar{x}^*$ for all moments of time $t \in [0, \Delta]$, each type $\theta$'s payoff vector remains constant $\mathbf{F}[\bar{x}^*](\theta)$. This implies that the switching rate of a type-$\theta$ agent from action $i$ to $j$ is unchanged from $Q_{ij}(F(\bar{x}^*), \theta)$ for $t \in [0, \Delta]$. Hence we can explicitly calculate the path $\{x_t(\theta) \,:\, t \in [0, \Delta]\}$ from $x_0(\theta)$. Fix a moment of time $T \in [0, \Delta)$ and express the path as the transition from time $T \in [0, \Delta]$. For each $\tau \in [-T, \Delta - T]$,

$$x_{a,T+\tau}(\theta) = \begin{cases} 1 - \sum_{i \in \mathcal{A} \setminus \{a\}} x_{a,T}(\theta) \exp\left(-Q_{ij}(F(\bar{x}^*), \theta)\tau\right) & \text{if } \theta \in \beta_a^{-1}(\bar{x}^*), \\ x_{a,T}(\theta) \exp\left(-Q_{aj}(F(\bar{x}^*), \theta(\omega))\tau\right) & \text{if } j \neq i \text{ and } \theta \in b_j^{-1}(\bar{x}^*). \end{cases}$$

The aggregate state is thus expressed as

$$\bar{x}_{a,T+\tau} = \int_{\beta_a^{-1}(\bar{x}^*)} \left\{ 1 - \sum_{i \in \mathcal{A} \setminus \{a\}} x_T(\theta) \exp\left(-Q_{ia}(F(\bar{x}^*), \theta)\tau\right) \right\} \mathbb{P}_\Theta(d\theta)$$

$$+ \sum_{j \in \mathcal{A} \setminus \{a\}} \int_{\beta_j^{-1}(\bar{x}^*)} x_{a,T}(\theta) \exp\left(-Q_{aj}(F(\bar{x}^*), \theta)\tau\right) \mathbb{P}_\Theta(d\theta)$$

$$= \mathbb{P}_\Theta(\beta_a^{-1}(\bar{x}^*)) - \int_{\beta_a^{-1}(\bar{x}^*)} \sum_{i \in \mathcal{A} \setminus \{a\}} \exp\left(-Q_{ia}(F(\bar{x}^*), \theta)\tau\right) x_T(d\theta)$$

$$+ \sum_{j \in \mathcal{A} \setminus \{a\}} \int_{b_j^{-1}(\bar{x}^*)} \exp\left(-Q_{aj}(F(\bar{x}^*), \theta)\tau\right) X_{a,T}(d\theta) \tag{A.2}$$

For each $a \in \mathcal{A}$, define functions $E_{a,T}^I, E_{a,T}^O : \mathbb{R} \to \mathbb{R}$: for each $\tau \in [-T, \Delta - T]$

$$E_{a,T}^I(\tau) := \int_{-\infty}^{\infty} \exp\left(-q_{a,T}^I \tau\right) \check{Q}_{a,T}^I(dq_{a,T}^I)$$

$$= \int_\Theta \sum_{i \in \mathcal{A} \setminus \{a\}} \exp\left(-Q_{ia}(F(\bar{x}^*), \theta)\tau\right) x_T(d\theta \cap \beta_a^{-1}(\bar{x}^*))$$

$$= \sum_{i \in \mathcal{A} \setminus \{a\}} \int_{\beta_a^{-1}(\bar{x}^*)} \exp\left(-Q_{ia}(F(\bar{x}^*), \theta)\tau\right) x_T(d\theta),$$

$$E_{a,T}^O(\tau) := \int_{-\infty}^{\infty} \exp\left(-q_{a,T}^O \tau\right) \check{Q}_{a,T}^O(dq_{a,T}^O)$$

$$= \int_\Theta \sum_{j \in \mathcal{A} \setminus \{a\}} \exp\left(-Q_{aj}(F(\bar{x}^*), \theta)\tau\right) X_{a,T}(d\theta \cap \beta_j^{-1}(\bar{x}^*))$$

$$= \sum_{j \in \mathcal{A} \setminus \{a\}} \int_{\beta_j^{-1}(\bar{x}^*)} \exp\left(-Q_{aj}(F(\bar{x}^*), \theta)\tau\right) X_{a,T}(d\theta).$$

With this expression, we can simplify (A.2) as

$$\bar{x}_{a,T+\tau} = \mathbb{P}_\Theta(\beta_a^{-1}(\bar{x}^*)) - E_{a,T}^I(\tau) + E_{a,T}^O(\tau).$$

The aggregate state staying at the aggregate equilibrium $\bar{x}_{a,T+\tau} = \bar{x}_a^* = \mathbb{P}_\Theta(\beta_a^{-1}(\bar{x}^*))$ implies

$$E_{a,T}^I(\tau) = E_{a,T}^O(\tau).$$

This holds for any $\tau \in [-T, \Delta - T)$. According to Lemma 1, this further implies that the distributions of the two random variables $q_{a,T}^I$ and $q_{a,T}^O$ coincide with each other at time $T$:

$$\check{Q}_{a,T}^I \equiv \check{Q}_{a,T}^O.$$

So, in particular at time $T = 0$, we obtain $\check{Q}_{a,0}^I \equiv \check{Q}_{a,0}^O$, i.e., the detailed balancing condition. $\square$

## A.5 Proof of Theorem 7

*Proof.* Consider the case in which $\check{Q}^O_{I,0}$ second-order stochastically dominates $\check{Q}^I_{I,0}$ and prove that there exists a solution trajectory of the tBRD with $\bar{x}_t < \bar{x}^*$ for all $t > 0$. The uniqueness of the solution path guarantees that this is the only solution trajectory and thus it must be the case that $\bar{x}_t < \bar{x}^*$ for all $t > 0$.

As we argued in the main text, second-order stochastic dominance of $\check{Q}^O_{I,0}$ over $\check{Q}^I_{I,0}$ first implies $\dot{\bar{x}}_0 < 0$ and thus guarantees that $\bar{x}_t < \bar{x}_0 = \bar{x}^*$ for small enough $t > 0$.

Consider an agent with type $\theta \geq F^0(\bar{x}^*)$, whose best response is O at time 0. Under positive externality, notice that $F^0(\bar{x}_t)$ is smaller than $F^0(\bar{x}^*)$ as long as $\bar{x}_t < \bar{x}^*$. Thus this agent should keep action O as the best response at time $t$. If such an agent has not switched to action O yet at such time $t$, the payoff deficit is $\check{F}[\bar{x}_t](\theta) = \theta - F^0(\bar{x}_t)$. Positive externality also implies that this payoff deficit at time $t$ cannot be smaller than that at time 0, $\check{F}[\bar{x}^*](\theta) = \theta - F^0(\bar{x}^*)$. As $Q_{IO}$ is assumed to be non-decreasing in the payoff deficit, this implies that the agent's switching rate is at least $Q_{IO}(F(\bar{x}^*), \theta)$. Therefore, transition of the Bayesian strategy for any type $\theta > F^0(\bar{x}^*)$ follows

$$\dot{x}_t(\theta) = -Q_{IO} \circ \mathbf{F}[\bar{x}_t](\theta) x_t(\theta) \leq -Q_{IO}(F(\bar{x}^*), \theta) x_t(\theta),$$

$$\therefore \quad x_t(\theta) \leq \exp\left(-Q_{IO}(F(\bar{x}^*), \theta) t\right) x_0(\theta).$$

On the other hand, positive externality and $\bar{x}_t < \bar{x}^*$ implies that, if an agent has action I as the best response at time $t$, so does it at time 0; the agent's type must be $\theta \leq F^0(\bar{x}_t) < F^0(\bar{x}^*)$. If the agent has not yet taken I at time $t$, the payoff deficit is $\check{F}[\bar{x}_t](\theta) = F^0(\bar{x}_t) - \theta$. This cannot be greater than $\check{F}[\bar{x}^*](\theta) = F^0(\bar{x}^*) - \theta$. By non-decreasingness of $Q_{OI}$, this implies that the agent's switching rate is at greatest $Q_{OI}(F(\bar{x}^*), \theta)$. Let $x_{O,t} = 1 - x_t$. Then, the transition of the Bayesian strategy for any type $\theta < F^0(\bar{x}^*)$ satisfies[15]

$$\dot{x}_{O,t}(\theta) \geq -Q_{OI} \circ \mathbf{F}[\bar{x}_t](\theta) x_{O,t}(\theta) \geq -Q_{OI}(F(\bar{x}^*), \theta) x_{O,t}(\theta),$$

$$\therefore \quad x_{O,t}(\theta) \geq \exp\left(-Q_{OI}(F(\bar{x}^*), \theta) t\right) x_{O,0}(\theta).$$

$$\text{i.e.} \quad x_t(\theta) \leq 1 - \exp\left(-Q_{OI}(F(\bar{x}^*), \theta) t\right) x_{O,0}(\theta).$$

It follows that[16]

$$\bar{x}_t \leq \int_{-\infty}^{F^0(\bar{x}^*)} \{1 - \exp\left(-Q_{OI}(F(\bar{x}^*), \theta) t\right) x_{O,0}(\theta)\} P_\Theta(d\theta)$$

$$+ \int_{F^0(\bar{x}^*)}^{+\infty} \exp\left(-Q_{IO}(F(\bar{x}^*), \theta) t\right) x_0(\theta) P_\Theta(d\theta)$$

$$= \bar{x}^* + \int_0^{+\infty} (-e^{-qt}) \check{Q}^I_{I,0}(dq) - \int_0^{+\infty} (-e^{-qt}) \check{Q}^O_{I,0}(dq).$$

For the equality, we used the assumption that $\bar{x}^*$ is an aggregate equilibrium with Assumption 2, i.e., $P_\Theta(F^0(\bar{x}^*)) = \bar{x}^*$. Function $-e^{-qt}$ is strictly increasing and strictly concave. Thus the second-order stochastic dominance of $\check{Q}_{I,0}^O$ over $\check{Q}_{I,0}^I$ implies that the latter integral in the above equation is greater than the former integral. Therefore, we have

$$\bar{x}_t < \bar{x}^*.$$

---

[15] Note that the best response for type $\theta < F^0(\bar{x}^*)$ may be switched to O by time $t$ and thus $\dot{x}_{O,t}(\theta)$ may be positive. But here we obtain a lower bound on $\dot{x}_{O,t}(\theta)$ and thus an upper bound on $x_t(\theta)$.

[16] Note that the upper bound on the last line is $\bar{\bar{x}}_t$, defined in (8).

The claim for the case in which $\check{Q}_{I,0}^O$ is stochastically dominated by $\check{Q}_{I,0}^I$ can be proved similarly. □

## A.6 Proof of Theorem 8

*Proof.* It is verified in the proof of Theorem 7 that $\bar{\bar{x}}_t \geq \bar{x}_t$. Thus, (9) implies $\bar{x}_t$ surely goes below the distributional critical mass $\bar{x}^\dagger$. Once the aggregate strategy hits $\bar{x}^\dagger$, it further decreases and cannot return above $\bar{x}^\dagger$ however the strategy composition changes. □